\documentclass[prc,reprint,aps,amsmath,amssymb,showpacs,nofootinbib,preprintnumbers,superscriptaddress]{revtex4-2}
\usepackage{graphicx}
\usepackage[pdftex]{hyperref}

\usepackage{cases}
\usepackage{pgf,tikz}
\usepackage{graphicx}
\usepackage{booktabs}

\usepackage{slashed} 

\usepackage{hyperref}
\usepackage[nameinlink]{cleveref}
\crefdefaultlabelformat{#2\textup{#1}#3}
\creflabelformat{equation}{#2\textup{#1}#3}

\newcommand{\ud}{\mathrm{d}}
\newcommand{\pd}{\partial}

\newcommand{\al}{\alpha}

\newcommand{\diff}[2]{\frac{\ud #1}{\ud #2}}

\newcommand{\lrp}[1]{\left(#1\right)} 
\newcommand{\lrb}[1]{\left[#1\right]} 

\newcommand{\MeV}{\text{MeV}}
\newcommand{\GeV}{\text{GeV}}

\newcommand{\msol}{\mathrm{M_\odot}}

\newcommand{\fmic}{\text{fm}^{-3}}

\newcommand{\const}{\text{const}}
\newcommand{\diag}{\text{diag}}
\newcommand{\tov}{\text{TOV}}
\newcommand{\pqcd}{\text{pQCD}}
\newcommand{\ceft}{\text{$\chi$EFT}}
\newcommand{\crit}{\mathrm{crit}}

\newcommand{\cqm}{\text{CQM}}

\newcommand{\dpmin}{\Delta P_\mathrm{min}}
\newcommand{\dpmax}{\Delta P_\mathrm{max}}
\newcommand{\cfl}{\mathrm{CFL}}

\newcommand{\cmax}{C_{s, \mathrm{max}}}
\newcommand{\cov}{\mathrm{Cov}}
\newcommand{\msbar}{\overline{\mathrm{MS}}}

\newcommand{\UCB}{Department of Physics, University of California Berkeley, Berkeley, CA 94720}

\begin{document}
\title{Reexamining constraints on neutron star properties from perturbative QCD}
\author{Dake Zhou}
\email{dkzhou@berkeley.edu}
\affiliation{Department of Physics, University of Washington, Seattle, WA 98195}
\affiliation{Institute for Nuclear Theory, University of Washington, Seattle, WA 98195}
\affiliation{\UCB}
\affiliation{Department of Physics and Astronomy, Northwestern University, Evanston, IL 60208}

\preprint{N3AS-23-026}

\date{\today}

\begin{abstract}
The implications of perturbative QCD (pQCD) calculations on neutron stars are carefully examined.
While pQCD calculations above baryon chemical potentials $\mu_B\simeq2.4$ GeV demonstrate the potential of ruling out a wide range of neutron star equations of state (EOSs), such constraints only affect the most massive neutron stars in the vicinity of the Tolman-Oppenheimer-Volkoff (TOV) limit, resulting in constraints that are orthogonal to current or expected astrophysical bounds.
In the most constraining scenario, pQCD considerations favor low values of the squared speed sound  $C_s$ at high $\mu_B$ relevant for the most massive neutron stars, but leave predictions of the radii and tidal deformabilities almost unchanged.
Such considerations become irrelevant if the maximum speed of sound squared inside neutron stars does not exceed about $\cmax\simeq0.5$, or if pQCD breaks down below $\mu_B\simeq2.9$ GeV. 
Furthermore, the large pQCD uncertainties preclude any meaningful bounds on the neutron star EOS at the moment. 
Interestingly, if pQCD predictions for the pressure at around $\mu_B\simeq2.5$ GeV are refined and found to be low ($\lesssim 1.5$ GeV/fm$^3$), evidence for a soft neutron star inner core EOS in combination with the existence of two-solar-mass pulsars would indicate the presence of color superconductivity beyond neutron star densities. 
I point out that two-solar-mass pulsars place robust upper bounds on this non-perturbative effect and require the pairing gap to be less than $\Delta_\cfl\leq500~\MeV$ at $\mu_B\simeq2.5~\GeV$.
\end{abstract}

\maketitle

\section{Introduction}

An {\em ab initio} QCD-based calculation of dense matter beyond a few times the nuclear saturation density ($n_0=0.16$ fm$^{-3}$) relevant for the interior of neutron stars remains challenging.
This strongly interacting regime not only precludes perturbative calculations but also discourages applications of lattice methods due to the sign problem~\cite{Troyer:2004ge, deForcrand:2009zkb, Kaplan:2009yg}. 
Such obstacles have deprived us of a clear understanding of neutron stars even though almost a century has passed since their conception by Landau and by Zwicky and Baade~\cite{Baade_1934b}.

In contrast to the stagnation in this strongly coupled region, significant progress has been made at the low- and ultra high-density frontiers. On one hand, for baryon number densities below about $\simeq 2n_0$, an effective description based on symmetries of QCD has proven fruitful in predicting the properties of neutron-rich matter. Dubbed chiral effective field theory (\ceft)~\cite{Weinberg:1968de,Weinberg:1990rz,Weinberg:1991um,Weinberg:1992yk,Kaplan:1996xu,Kaplan:1998tg,Kaplan:1998we,Beane:2001bc}, it allows for and arranges all possible operators respecting chiral symmetry through a power counting scheme~\cite{Weinberg:1992yk, Kaplan:1998tg}.
The result is a systematic expansion in nucleon momenta that not only enables calculations of the equations of state but also provides error estimates through order-by-order comparisons~\cite{Kruger:2013kua,Wlazlowski:2014a,Gandolfi:2014,Lynn:2015,Drischler:2017wtt,Drischler:2020yad}.
For example, the recent next-to-next-to-next leading order (N3LO) calculation reveals that the $1\sigma$ uncertainties associated with the pressure of pure neutron matter (PNM) to be $\approx 10\%$ at $n_B=n_0$, and worsen to about $40\%$ at $2n_0$ ~\cite{Drischler:2020yad}.
On the other hand, at densities above $\simeq 30-50 n_0$, asymptotic freedom of QCD suggests perturbative treatments become valid and useful. The pioneering work of \cite{Freedman:1976xs,Freedman:1976ub} confirmed that cold quark matter resembles closely to a Fermi gas in this region, and has promoted significant efforts in simplifying and improving calculations for cold and dense QCD~\cite{Manuel:1995td,Kurkela:2009gj,Gorda:2021kme,Kurkela:2016was,Gorda:2018gpy}.

Given the scarcity of available information directly related to the phase of matter at supranuclear densities in between, gleaning as many insights as possible from both ends is a desirable undertaking.
That high-density pQCD calculations may place limits on NS EOSs through thermodynamic consistency requirements is well-known, and
there is only a potential for constraints because the current uncertainties associated with pQCD predictions are simply too large to place meaningful bounds.
However, a renewed wave of interest in this topic is recently generated by \cite{Komoltsev:2021jzg}.
That letter along with the subsequent papers~\cite{Gorda:2022jvk} including attempts to clarify it~\cite{Somasundaram:2022ztm} implicitly made several yet to be justified assumptions and have led to apparently widespread misunderstandings in the literature. 

In this first paper of the series exploring the interplay between high- and low-density theoretical inputs I focus on the implications of pQCD calculations on neutron stars and the underlying EOS.
One objective is to clarify the misunderstandings on this subject. Specifically, it will be shown that due to the undetermined strength of non-perturbative effects only one type of the bounds derived from thermodynamic relations may be viewed as a constraint on NS EOSs.
Furthermore, while \pqcd~ constraints do demonstrate the potential of ruling out a considerable region in the EOS space, astrophysical observables are barely affected by these high-density inputs, as bounds from pQCD are mostly orthogonal to those from astrophysics.

The rest of the paper is organized as follows. A simple parameterization of \ceft~ EOSs is discussed in \cref{sec:nseos},
and the pQCD EOS for cold quark matter is reviewed in \cref{sec:pqcd}. In \cref{sec:thermo} I review the previously-known model-independent bounds on the pressure that the low- and high-density EOSs must satisfy~\cite{Rhoades:1974fn,Koranda:1996jm,Lattimer:2000nx,Drischler:2020fvz}.
In \cref{sec:ns} I apply these bounds and carefully examine the underlying assumptions, the applicable ranges, and their consequences.
Main results and future directions are summarized in \cref{sec:concl}.
Throughout the discussion I adopt the natural unit system in which $G=\hbar=c=1$.

\section{neutron star EOS}\label{sec:nseos}

In this section I give an overview of the parameterization of neutron star EOSs used in this work. Some details are reported in ~\cite{Forbes:2019xaz,Zhou:2024usp}.

\subsection{parameterizing \ceft-based EOSs with correlated uncertainties}

Recently, refs \cite{Melendez:2019izc, Drischler:2020yad} studied truncation errors of \ceft~ based on the assumptions that its predictions for energy per particle admit a polynomial expansion in momenta, and that the unknown coefficients associated with higher order terms are natural in size. 
Using a Gaussian Process (GP) interpolant the authors inferred the size and characteristic length scales of the inter-density correlations of the polynomial coefficients, and used these to obtain extrapolated truncation errors. The resulting correlated uncertainties are encoded in the covariance matrix $\cov(E_i, E_j)$ where $E_{i,j}$ are energies per particle at the ith and jth tabulated density points, excluding the rest mass. Although one can reconstruct the GP interpolant reported there and use it to generate EOS samples, this approach may not be easily extended to construct beta-equilibrium EOSs that respect the underlying inter-density correlations.

Here, I present a minimal, faithful parameterization of the \ceft~ calculations that captures not only the central values but also the correlated truncation errors. 
The type of correlation I focus on is the inter-density correlation, as it is the most significant one in approach to describe the beta-equilibrium neutron-rich matter, to be discussed below. For a discussion on the correlation between pure neutron matter (PNM) and symmetric nuclear matter (SNM) EOSs, see ~\cite{Drischler:2020yad,Drischler:2020fvz}.
At the core of this parameterization lies the eigenvalue decomposition (singular value decomposition) of the covariance matrix 
\begin{equation}\label{eq:ceft_cov}
U S U^T=||\cov(E_i, E_j)||.
\end{equation}
The eigenvectors in $U$ are orthonormal since the covariance is symmetric.
In particular, each eigenvector $U_i$ represents a correlation among different densities of the truncation error, with the $1\sigma$ variance given by the square root of its corresponding eigenvalue $\sqrt{\diag(S)_i}$.
For both the chiral potentials with $450$ MeV and $500$ MeV cutoffs reported in \cite{Drischler:2020yad}, the largest (second largest) entry in $\sqrt{\diag(S)}$ accounts for about $90\%$ ($8\%$) of the variances encoded in the full covariance matrix, 
suggesting their corresponding eigenvectors, which shall be referred to as the most significant eigenvectors, are adequate for current and future astrophysical applications. 
This procedure is suitable for both PNM and SNM calculations, and for concreteness let us focus on the PNM EOS of the form $E(n_n)$ where $n_n$ is the neutron number density.

The two most significant eigenvectors form a set of orthogonal basis to describe the correlated uncertainties. 
Denote $E_0(n_n)$ the central values of energy per particle predicted by \ceft, the proposed parameterization that incorporates correlated uncertainties takes the form
\begin{multline}\label{eq:eft_eos}
\widehat E(n_n) = {\widehat E}_0(n_n) + \sum _{i=1}^{2} a_i  {\widehat E}_i(n_n) + m_n,\\
 \quad\text{with } {\widehat E}_i(n_n)=\sqrt{\diag(S)_i}\ U_i(n_n).
\end{multline}
Above, $m_n$ is the mass of neutron. The variance associated with each eigenvector is absorbed into the definition of basis functions ${\widehat E}_i(n_n)$, so that the (uncorrelated) coefficients $a_i$ are  
independent identical draws from the standard normal distribution $\mathcal{N}(0,1)$. 
This procedure effectively projects out the high-dimensional covariance matrix to a lower dimension subspace spanned by the basis functions ${\widehat E}_{i=1,2}(n_n)$.

\begin{figure}
	\includegraphics[width=0.98\linewidth]{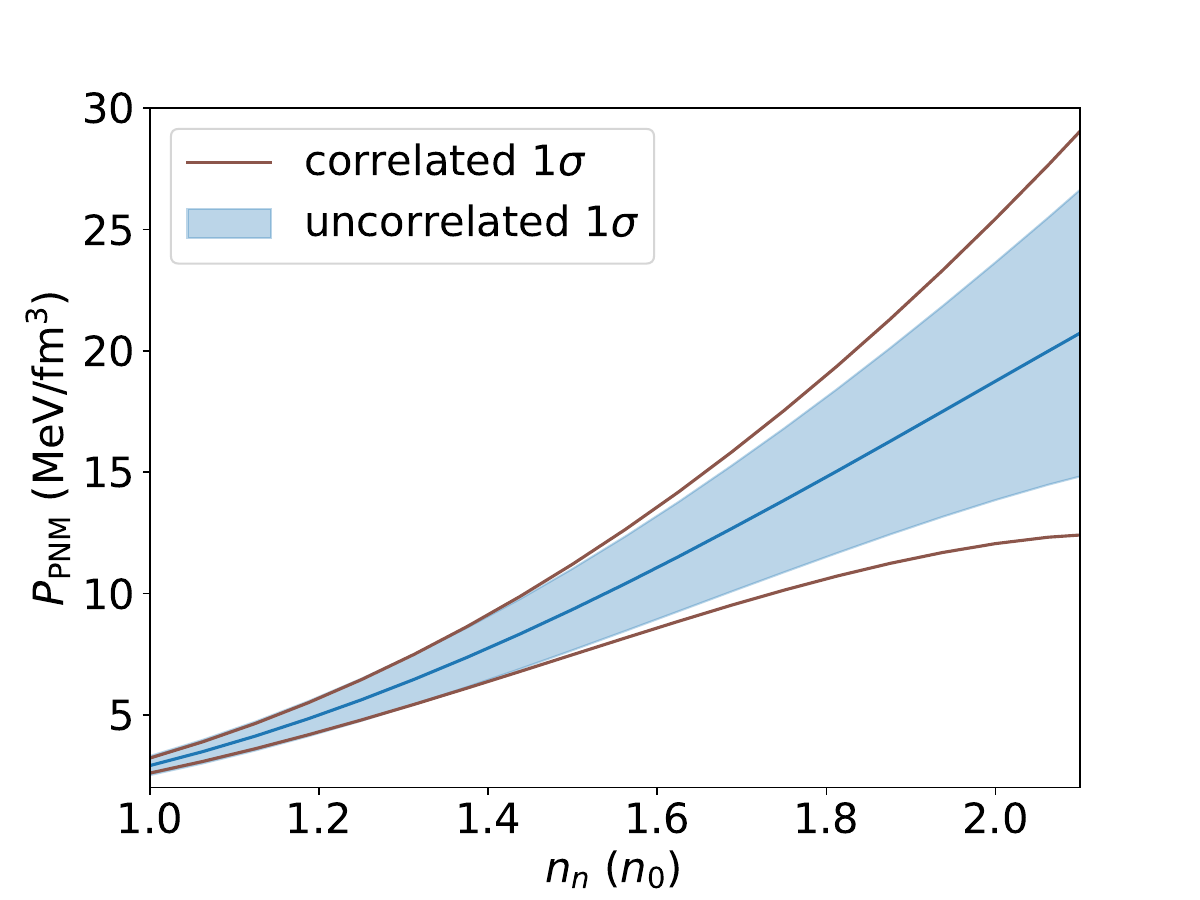}
	\caption{ The pressure of the PNM along with $1\sigma$ uncertainties from \ceft~ calculations at N3LO.
	This EOS is based on the chiral potential with a $500$ MeV cutoff reported in~ \cite{Drischler:2020yad}.
	The correlated truncation errors are obtained according to \cref{eq:ceft_cov} and \cref{eq:eft_eos}, and are shown as the brown lines at the $1\sigma$ level.
	 Uncorrelated uncertainties (blue band) underestimate the error in the pressure by as much as $30\%$ near $n_B=2n_0$,
	even though it agree with the correlated version on energy per particle within $10\%$ (not shown here).
	}\label{fig:n3lo500P}
\end{figure}

These basis functions, along with the central values $E_0(n_n)$, are well-behaved and can be faithfully represented by polynomials. For optimal fittings, polynomials of degree $\gtrsim5$ are sufficient as the fitted EOS as well as derivatives are stationary with increasing polynomial degrees. 
This allows us to turn the discrete tabulated EOS from \ceft~ calculations to a compact, faithful, and continuous representation \cref{eq:eft_eos}.
The resulting EOSs in this parameterization are shown in \cref{fig:n3lo500P}. 
The pressure follows from the thermodynamic relation $P=n_n^2(\ud E/\ud n_n)$, and the $1\sigma$ upper and lower bounds of correlated errors are obtained by setting $a_1=a_2=\pm1$ in \cref{eq:eft_eos}.
For comparison, the uncorrelated $1\sigma$ errors obtained by taking the square root of $\diag(||\cov(E_i, E_j)||)$ are shown as the blue band.
The importance of inter-density correlations is evident as the boundaries of the blue band underestimate the error in pressure by almost $1/3$ near $n_B=2n_0$.
Inter-density correlations are positive since the energy at a given point is less likely to be low if its neighboring points are predicted to have high energies by the chiral interaction.

The EOS in \cref{eq:eft_eos} can be parameterized either against number densities $n_n$ or Fermi momenta $k_F$. 
The covariance matrix obtained in \cite{Drischler:2020yad} in fact involves $E(k_F)$ tabulated on an equally spaced grid in Fermi momenta $k_F$, as inter-density correlations appear more natural on this grid.
It is not straightforward to translate this GP interpolated EOS on the uniform grid of $k_F$ to another GP representation on a grid of different thermodynamic variables, as the physical inter-density correlations are not preserved in the process.
This is a limitation due to GP where the kernels are translational invariant.
Here, choosing either $k_F$ or $n_n$ is fine since the correlations captured by the basis functions are invariant under change of variables. 
Interpolation by polynomials does introduce systematics, though a simple estimate by dropping half the points and reproducing them using the remaining data points seems to indicate they are controlled and no more than $\approx5\%$.

This prescription works well for both PNM and SNM EOSs, but the approach focuses on the PNM calculation from \cite{Drischler:2017wtt, Drischler:2020yad}. 
The EOS for nuclear matter in beta-equilibrium is then constructed based on an expansion in proton fraction $x_p=n_p/n_B$ described in ~\cite{Forbes:2019xaz}.
This approach incorporates nuclear saturation properties by imposing a boundary condition at $x_p=1/2$. Unlike those based on the parabolic expansion centered around $x_p=1/2$, the resulting beta-equilibrium EOS is only moderately sensitive to the properties of SNM.
While theoretical and experimental probes of SNM near and below saturation densities are very valuable inputs, they may suffer from currently unknown systematic uncertainties, as the recent neutron skin measurements might have indicated~\cite{PREX:2021umo,Reed:2021nqk}. 
Adding that neutron stars are mostly neutrons, this informed yet flexible  approach is a noteworthy alternative to those in the literature for connecting nuclear physics to neutron star observables.
For the purpose of this work, the results are insensitive to the choices between SNM-centered or PNM-centered expansions, and only results based on the latter will be presented.
A systematic comparison of the two and careful examinations of low-energy nuclear inputs will be reported in another work focusing on low-density EOSs.

\subsection{inner core EOS}

At higher densities above $n_{\ceft}\simeq1-2n_0$ two approaches to describe the NS inner core are adopted.
Firstly, I employ the limiting EOSs that have been long known in the literature.
Their predictions in the mass-radius plane encircle all physical possibilities as they yield 
lower and upper bounds on the pressure at given densities.
Due to their ability to delimit the physical regions in the EOS space, they lie at the center of the model-independent framework that propagates the information of cold quark matter at ultra-high densities down below.
They will be discussed in detail in \cref{sec:thermo}.
By examining how pQCD calculations impact these extreme scenarios I demonstrate that pQCD cannot improve bounds on NS static observables in \cref{sec:ns}.

In the second approach, I randomly generate inner core EOSs to explore how pQCD impacts the statistics of NS observables.
Since pQCD does not change the boundaries in the mass-radius plane, their impacts on the probabilistic distributions (if and when relevant) would dependent upon the assumed NS models.
Carefully examining the underlying assumptions in the EOS parameterizations is thus important to ensure robust constraints.
I employ a family of speed of sound parameterizations detailed in~\cite{Zhou:2024usp}. Several subsets have been reported elsewhere in  e.g.~\cite{Tews:2018kmu,Landry:2018prl,Annala:2019puf}.
I have checked that the results below are robust against different choices of priors on EOS parameters (implicit or explicit), and across different subsets of inner core parameterizations.

\section{perturbative QCD}\label{sec:pqcd}

\begin{figure}
	\includegraphics[width=0.98\linewidth]{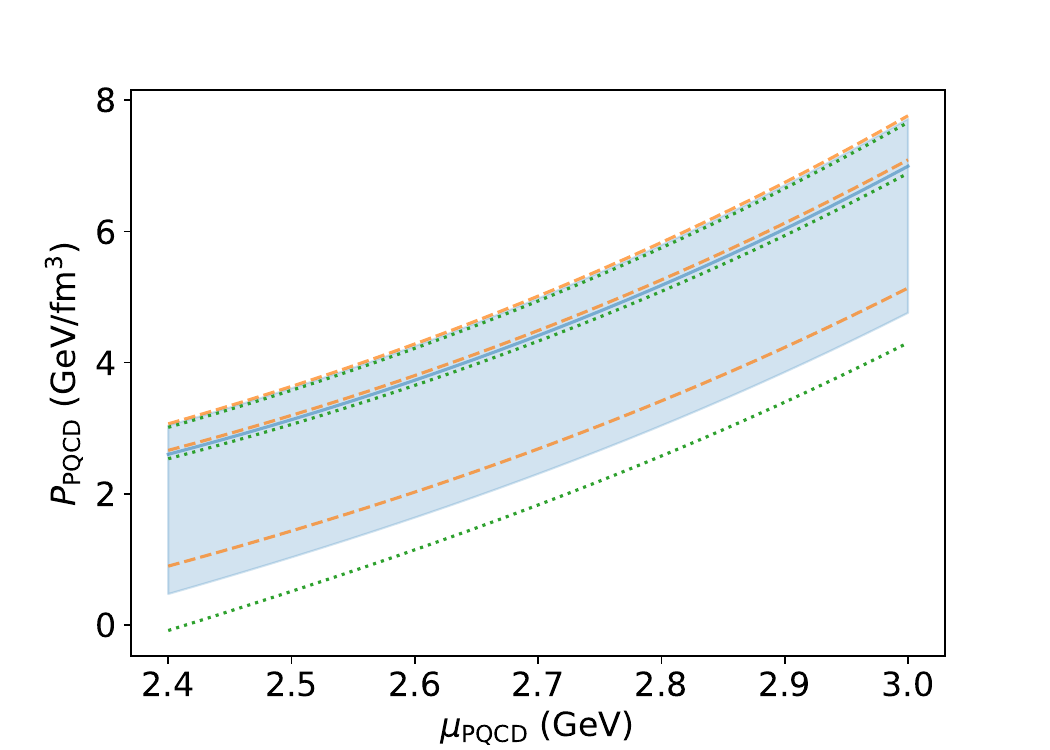}
	\caption{The pQCD EOS up to N2LO. The blue band shows the renormalization scale uncertainties assuming $\alpha_s(2~\GeV)=0.2994$, where the lower boundary corresponds to $X=1$, the upper boundary corresponds to $X=4$, and the blue solid line in between to $X=2$. The green (orange) lines mark the same ranges for $X=[1,2,4]$, but assume the $0.5\sigma$ upper (lower) bound of $\alpha_s(2~\GeV)$ (see \cref{sec:alphas}). 
	}\label{fig:pqcd}
\end{figure}

At asymptotic densities, QCD becomes perturbative and renders useful an expansion in the strong coupling $\alpha_s$. The calculation of cold quark matter EOS up to N2LO is first performed in \cite{Freedman:1976xs,Freedman:1976ub}  in the momentum space subtraction renormalization scheme, and remains the state-of-art full order calculation.
Subsequent works formulated it in the widely adopted $\msbar$ scheme\cite{Vuorinen:2003fs, Kurkela:2009gj}, in which the pressure of unpaired quark matter with $N_f=3$ massless flavors is given by
\begin{multline}\label{eq:pqcd}
P=N_f\frac{ (\mu_B/3)^4}{4\pi^2} 
\left\{1-\frac{2}{\pi}\al_s - \frac{N_f}{\pi^2}\al_s^2\log{\al_s} \right. \\
\left. -\frac{\al_s^2}{\pi^2} \lrb{{c_1+N_f \lrp{\log\frac{N_f}{\pi}-c_2}}
	+ \lrp{11-\frac{2}{3}N_f} \log{X}} \right.\\
\left.	+ \mathcal{O}(\alpha_s^3) \right\}
\end{multline}
Above, $c_1=18-11\log{2}$ and $c_2=0.535832\ldots$ are numerical constants. The renormalization scale $\bar\Lambda$ is implicit in $\alpha_s\equiv\alpha_s(\bar\Lambda)$ and is conveniently parameterized in terms of the baryon chemical potential by $X=\bar\Lambda/(\mu_B/3)$. 
Since the quark chemical potential $\mu_q=\mu_B/3$ is the characteristic scale of the system, one might expect a good choice of the renormalization scale to be close to it ($X\sim 1$).  
In ref \cite{Ipp:2003jy} the authors compared predictions of \cref{eq:pqcd} to an exactly solvable large $N_f$ limit where $X$ is varied between $1/2$ and $2$. It was found that the large $N_f$ prediction lies somewhere in $X\in[1,2]$, and is a bit closer to $X=2$.
Based on this observation, as well as empirical evidence from hot pQCD calculations, the dense quark matter community appears to have adopted fiducial values $X=1,2,4$ to estimate the renormalization scale uncertainties~\cite{Kurkela:2009gj}.
This range is shown in \cref{fig:pqcd} as the blue band.

Note that the appearance of $X$ in \cref{eq:pqcd} is an artifact of the perturbation theory since physical observables shall be independent of it. Once higher order corrections are included, pQCD predictions are expected to show diminished sensitivity towards the renormalization scale,  assuming pQCD is converging at given $\mu_B$.
In this sense, the uncertainty associated with $X$ is in fact a truncation error of the perturbation series.
Efforts estimating the truncation error based on naturalness considerations are underway and will be reported in future work.

Another often neglected source of uncertainties comes from the running of $\alpha_s$. 
In the main text, the two-loop solution of $\alpha_s$ consistent with \cref{eq:pqcd} is used.
The uncertainty associated with the inferred value of $\alpha_s$ can have a sizable impact on pQCD predictions. This is shown in \cref{fig:pqcd} as the dashed and dotted lines.
The effect is especially prominent towards lower values of $X$, where a percentage difference in the reference value $\alpha_s(2~\GeV)$ would shift the pressure on the order of $20-50\%$.
Note that I do not claim the choice adopted here is the best.
A few other options are discussed in \cref{sec:alphas}.
I merely wish to point out that the uncertainties associated with $\alpha_s$ have to be properly accounted for before claiming robust constraints, and that extracting $\al_s$ in the infrared is inherently challenging. For a recent review see~\cite{Deur:2016tte}.

Recently, partial N3LO contributions are computed in \cite{Gorda:2021kme,Gorda:2021znl,Gorda:2018gpy,Gorda:2023mkk} for the soft and mixed gluon sectors by resumming diagrams using EFT techniques~\cite{Frenkel:1989br,Braaten:1989mz,Taylor:1990ia}. The newly obtained contributions are the first two positive terms in the series, and hence predict higher pressure at given chemical potentials. However, if calculations of cold dense QED~\cite{Gorda:2022zyc} or hot QCD~\cite{Braaten:1995jr,Braaten:1995cm} were to be of any guide, the unknown hard contribution at N3LO is likely negative, and the net contribution at full order N3LO may drive the pressure even lower than that of \cref{eq:pqcd}. In light of this, I take the cautious approach and employ the N2LO quark matter EOS in the main text. I comment on the impact of the soft contribution at N3LO and present the results in \cref{sec:n3lo}.

\section{The limiting EOSs and bounds on the pressure}\label{sec:thermo}

In this section, I review the consistency requirement on low- and high-density EOSs at zero-temperature implied by the limiting scenarios well-known in the literature~\cite{Rhoades:1974fn,Koranda:1996jm,Lattimer:2000nx,Drischler:2020fvz}.
These extreme scenarios are simply consequences of causality and stability of dense matter, and are often colloquially referred to as the maximally stiff and maximally soft EOSs.
They respectively yield the highest (lowest) and the lowest (highest) increments in pressure over arbitrary intervals in chemical potential (density).
Since the baryon chemical potential $\mu_B$ is the natural parameter in pQCD calculations, I will present these limiting EOSs in terms of $\mu_B$ in a pedagogical approach. 

In the $\mu_B-P$ plane, the pressure $P$ as a function of the baryon chemical potential $\mu_B$ must be continuous and differentiable. Furthermore, $P(\mu_B)$ is a convex function since
\begin{align}
\ud P/\ud \mu_B&=n_B,\label{eq:dpdmu}\\
\frac{\ud\log n_B}{\ud \log\mu_B}&=C_s^{-1}, \quad 0\leq C_s \leq1. \label{eq:cs}
\end{align}
where $C_s$ is the speed of sound squared, a non-negative quantity for stable phases. Causality also imposes the additional requirement that $C_s\leq1$. 
Together these relations suggest that for a given interval $\Delta \mu_B=\mu_H-\mu_L$ between two arbitrary points labeled ``L'' and ``H'', 
there exists an upper and a lower bound on $\Delta P=P_H-P_L$, where $P_H=P(\mu_H)$ and $P_L=P(\mu_L)$.

\begin{figure}
	\includegraphics[width=0.8\linewidth]{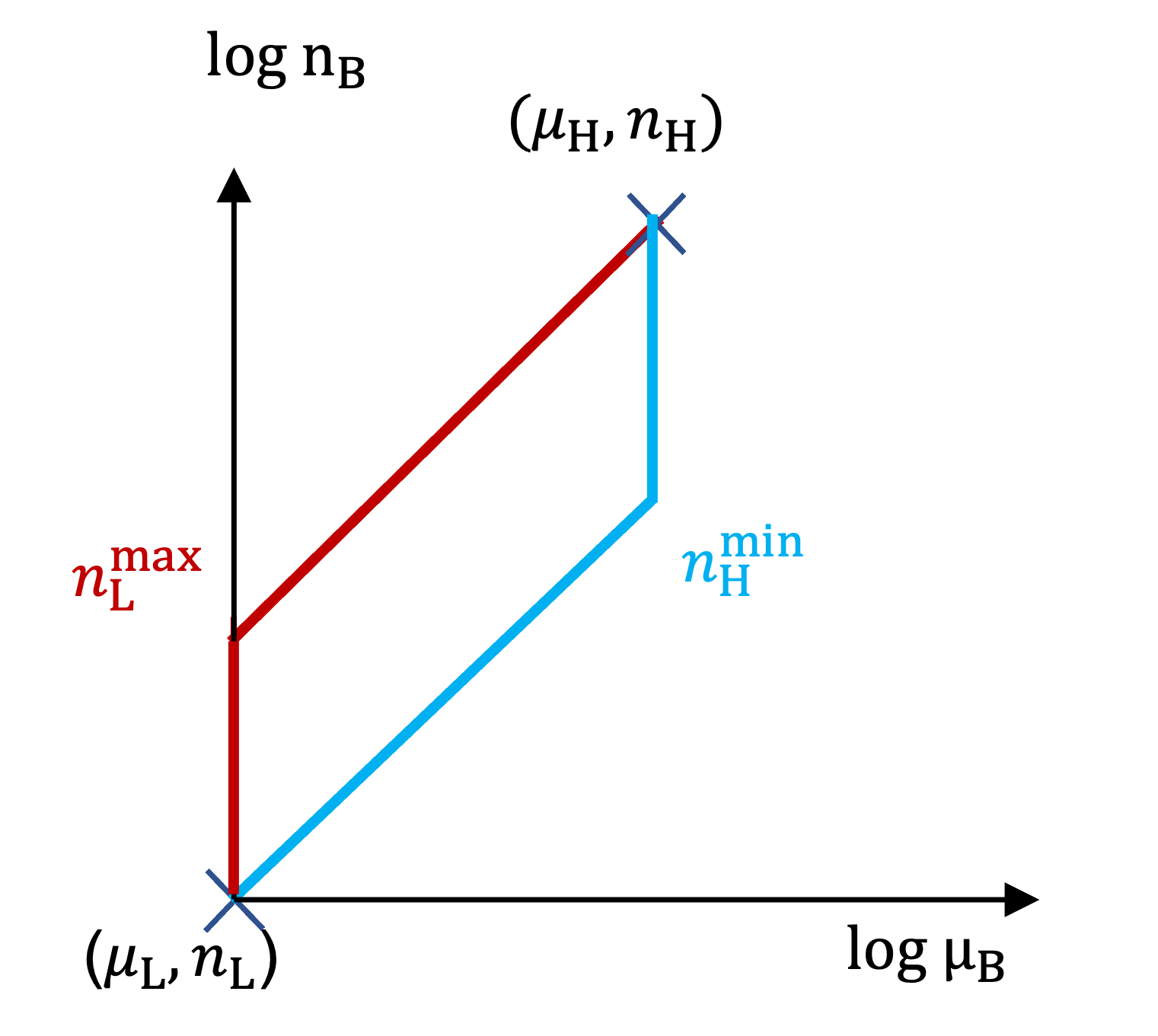}
	\caption{The maximally stiff (blue) and maximally soft (red) EOSs between $\mu_L$ and $\mu_H$ that yield $\Delta P_\mathrm{min}$ and $\Delta P_\mathrm{max}$.}\label{fig:construction}
\end{figure}

To obtain the upper bound on $\Delta P$, the largest possible derivative $\diff{P}{\mu_B}=n_B$ throughout the range $[\mu_L, \mu_H]$ is desired. 
The schematics for the limiting cases of $n_B(\mu_B)$ is shown in \cref{fig:construction}.
At $\mu_L$, in order to get the largest value of the number density, one may introduce a first order phase transition which can raise $n_B$ by an arbitrary amount at the expense of zero increase in $\mu_B$.
But since the slope in the $\log n_B-\log\mu_B$ plane is the inverse of the speed of sound squared, which must be no less than 1 as demanded by causality, jumps in the number density that are too large would overshoot $n_H$ at $\mu_H$.
To find the maximum of this value of $n_L$, one may start from $(\mu_H, n_H)$ and follow the line with the least possible slope $C_s^{-1}=1$ down to $\mu_L$, where it gives
\begin{equation}\label{eq:nLmax}
n_L^\mathrm{max} = n_H \exp\lrb{\int^{\log\mu_L}_{\log\mu_H} \ud\log\mu_B} = n_H \frac{\mu_L}{\mu_H}.
\end{equation}
This point is labeled in red in \cref{fig:construction}. 
From there, the EOS I just followed from $n_H$ downwards is the only allowed path to reach $(\mu_H, n_H)$ without violating causality $C_s\leq1$.
In fact, this EOS predicts the largest possible number densities between $\mu_L$ and $\mu_H$. 
Therefore, this construction shown in red in \cref{fig:construction} leads to the largest possible $\Delta P$ given by
$$	
\Delta P_\mathrm{max}=\frac{n_H}{\mu_H} \int^{\mu_H}_{\mu_L} \ud\mu_B \mu_B =\frac{n_H \mu_H}{2}\lrb{1-\lrp{\frac{\mu_L}{\mu_H}}^2}.
$$

Following similar arguments one finds that the EOS depicted in blue in \cref{fig:construction} bears the lowest number densities between $\mu_L$ and $\mu_H$.
It starts at $\mu_L$ with a constant $C_s=1$ and reaches $\mu_H$ with a number density 
\begin{equation}\label{eq:nHmin}
n_H^\mathrm{min} = n_L \exp\lrb{\int_{\log\mu_L}^{\log\mu_H} \ud\log\mu_B} = n_L \frac{\mu_H}{\mu_L},
\end{equation}
then climbs to $n_H$ via a first order phase transition. The resulting lower bound on $\Delta P$ reads
$$	
\Delta P_\mathrm{min}=\frac{n_L}{\mu_L} \int^{\mu_H}_{\mu_L} \ud\mu_B \mu_B=\frac{n_L \mu_L}{2}\lrb{\lrp{\frac{\mu_H}{\mu_L}}^2-1}.
$$

The constructions shown in red and blue in \cref{fig:construction} that give $\Delta P_\mathrm{max}$ and $\Delta P_\mathrm{min}$ are commonly referred to as the maximally soft and maximally stiff EOSs in the literature~\cite{Rhoades:1974fn,Koranda:1996jm,Lattimer:2000nx,Drischler:2020fvz}. 
When expressed in terms of number densities, which is useful for describing NS inner cores, they are specified by
\begin{align}
C_s(n_B)&=
\begin{cases}
0, \quad n_B\leq n_L+\Delta n_\mathrm{PT}\\
1, \quad n_B> n_L+\Delta n_\mathrm{PT}
\end{cases}
 &\mathrm{maximally\ soft}\label{eq:soft},\\
 C_s(n_B)&=
 \begin{cases}
 1, \quad n_B\leq n_L+\Delta n_\mathrm{onset}\\
 0, \quad n_B> n_L+\Delta n_\mathrm{onset}
 \end{cases}
  &\mathrm{maximally\ stiff}\label{eq:stiff}.
\end{align} 
Above, the parameters $\Delta n_\mathrm{onset}$ and $\Delta n_\mathrm{PT}$ that control the location and strength of phase transitions (the segment with $C_s=0$) are determined by the EOS at the endpoints, and I have just computed them indirectly in \cref{eq:nLmax,eq:nHmin}.
When the maximally stiff and soft constructions are used to bound NS masses and sizes, these parameters are fixed by the NS maximum mass, also known as the TOV limit $M_\tov$~\cite{Oppenheimer:1939ne,Tolman:1939jz}.
For specified $M_\tov$, the maximally stiff (soft) construction leads to the largest (smallest) radii at any given $M\leq M_\tov$.

To understand how $\dpmin$ and $\dpmax$ impose consistency requirements on the low-density endpoint when the high-density EOS at $H$ is known, I consider two opposite scenarios in which they separately obtain.
For low-density EOSs that are soft, i.e., predict high values of $P_L$ at given $\mu_L$, 
stiff EOSs in the region between $\mu_L$ and $\mu_H$ are needed to reach the given high-density point specified by $(\mu_H, n_H,  p_H)$.
If however the low-density EOS is way too soft such that $\Delta P=P_H-P_L < \Delta P_\mathrm{min}$, even the maximally stiff EOS between $\mu_L$ and $\mu_H$ would overshoot $p_H$ at $\mu_H$, suggesting a lack of causal and stable connections between the endpoints. 
Likewise, in the opposite limit, if the low-density EOS is too stiff predicting pressures that are hopelessly low, even the maximally soft EOS in the region $[\mu_L, \mu_H]$ (which yields $\Delta P_\mathrm{max}$) is unable to bring the pressure to $P_H$ at $\mu_H$.
Thus, for a given-high density endpoint $H$, $\dpmin$ and $\dpmax$ can in principle rule out low-density endpoints that are inconsistent with thermodynamics and requires
$$
\Delta P_\mathrm{min} \leq \Delta P\equiv P_H-P_L\leq \Delta P_\mathrm{max}
$$ 
without explicitly parameterizing the EOS over the range $[\mu_L, \mu_H]$.

Below, I will take the pQCD predictions at $\mu_\pqcd\equiv \mu_H\simeq 2.5-3$ GeV as the high-density boundary condition and study the implications of $\Delta P_\mathrm{max}$ and $\Delta P_\mathrm{min}$
\begin{align}
	\Delta P_\mathrm{max}&=\frac{n_\pqcd \mu_\pqcd}{2}\lrb{1-\lrp{\frac{\mu_L}{\mu_\pqcd}}^2}, \label{eq:dpmax} \\
	\Delta P_\mathrm{min}&=\frac{n_L \mu_L}{2}\lrb{\lrp{\frac{\mu_\pqcd}{\mu_L}}^2-1}, \label{eq:dpmin} 
\end{align}
on neutron stars.
While in principle both \cref{eq:dpmax,eq:dpmin} can incorporate perturbative information about the cold quark matter at $n_B\simeq30-50n_0$ to put limits on the NS EOS at lower densities, I shall demonstrate that only \cref{eq:dpmax} can be robustly considered as a constraint, 
given our current understanding of the zero-temperature QCD phase diagram.

\section{Implications for neutron stars}\label{sec:ns}

Because of the large separation in density scales between the perturbative regime where pQCD is applicable and the strongly-interacting phase relevant for the interior of neutron stars, it is conceivable that pQCD may only impact (if at all) rather extreme NS EOSs that predict high densities in NSs.
I shall show that such speculations are largely correct given the existence of two-solar-mass pulsars, but will clarify the context and criterion for the notion of ``closeness''.
I will also point out a necessary condition underlying all NS EOSs that are ``close enough'' to pQCD for which the latter could be relevant.
Since the central density of the most massive neutron star is the highest realized in nature thus closest to that of pQCD, I shall always attempt the matching of NS EOS to pQCD via \cref{eq:dpmax,eq:dpmin} at this density to maximize the constraining potentials, though occasionally matching at lower densities will be carried out for clarification purposes.
For clarity, I colloquially refer to the NS EOS at the center of maximum-mass stars as the TOV point.

Refs ~\cite{Komoltsev:2021jzg,Gorda:2022jvk} appear to be the first to report pQCD constraints on the NS EOS through \cref{eq:dpmax,eq:dpmin}.
The authors chose a few benchmark values of $n_L\simeq 4-8 n_0$ as the low-density matching point.
Ref~\cite{Somasundaram:2022ztm} pointed out matching at the TOV point is a better choice since the segment above it is not directly probed by NSs.
Below, I will highlight the need to choose the EOS-dependent $n_\tov$ and $\mu_\tov$, the number density and baryon chemical potential at the center of maximum-mass stars, as the low-density matching point, because they correlate with NS observables especially $M_\tov$.
Properly taking into account these correlations is crucial to minimize assumptions about the ultra-dense matter not realized in nature.

Astrophysical observations of NSs have already provide valuable insights into the EOS and need to be incorporated to describe realistic neutron stars.
In the main text, I will only explicitly take into account the existence of two-solar-mass NSs \cite{Demorest:2010bx,Antoniadis:2013pzd,NANOGrav:2019jur,Romani:2021xmb,Fonseca:2021wxt}, because (1) these measurements are accurate and robust as they exploit relatively clean systems and make minimal assumptions (2) their implied limits on $M_\tov$ can be translated to bounds on the NS EOS and NS radii through the maximally stiff and maximally soft EOSs, e.g.~\cite{Drischler:2020fvz,Drischler:2021bup}.
Since potential pQCD constraints  on NSs via $\dpmin$ and $\dpmax$ directly act upon the EOS space, this second benefit allows us to obtain model-independent bounds on NS static observables from pQCD.
Temporarily deferring astrophysical inputs on the NS radius $R$ and deformability $\Lambda$  also enables one to isolate the impact of high-density inputs, which helps elucidate the extent and strength of possible constraints on NSs coming from pQCD.
I will show in the main text that pQCD cannot appreciably affect NS global observables such as the radius and the tidal deformability before any astrophysical constraints on these observables are imposed, though it does have the potential of ruling out a considerable number of NS EOSs.
In \cref{sec:astro} I demonstrate the results remain valid when astrophysical bounds on $R$ and $\Lambda$ are taken into account.

The implications of $\dpmax$ and $\dpmin$ will be discussed separately due to their distinct dependencies on the endpoints and different sensitivities towards non-perturbative effects.
As mentioned in \cref{sec:nseos}, I employ the $\ceft$-based EOS up to $n_\ceft\simeq 1-2n_0$, where it is switched to either (1) the maximally stiff and soft inner cores to obtain bounds on NS observables; or (2) randomly generated inner cores to examine the statistics.

\subsection{$\Delta P_\mathrm{min}$ is not a bound on NSs (yet)}\label{sec:dpmin}

I begin by discussing the implications of $\Delta P_\mathrm{min}$. 
For a given low-density matching point $\mu_\tov$, $n_\tov$, and $P_\tov$, \cref{eq:dpmin} is solely a function of $\mu_\pqcd$.
It is thus convenient to propagate the information of the TOV point to higher densities by extending the EOS upward using $\dpmin$. 
I then compare the extrapolated pressure based on $\dpmin$ to pQCD predictions at $\mu_\pqcd$ in the perturbative regime.
This leads to a lower limit on $P_\pqcd\equiv P(\mu_\pqcd)$, requiring that
\begin{equation}\label{eq:pminbound}
P(\mu_\pqcd) \geq P_\tov+\Delta P_\mathrm{min}(\mu_\pqcd).
\end{equation}
As explained earlier, violations of \cref{eq:pminbound} would occur if $P_\tov$ is too high at $\mu_\tov$, so that $P_\tov+\Delta P_\mathrm{min}$ overshoots $P_\pqcd$ at the high-density matching point. 

\begin{figure}
	\includegraphics[width=\linewidth]{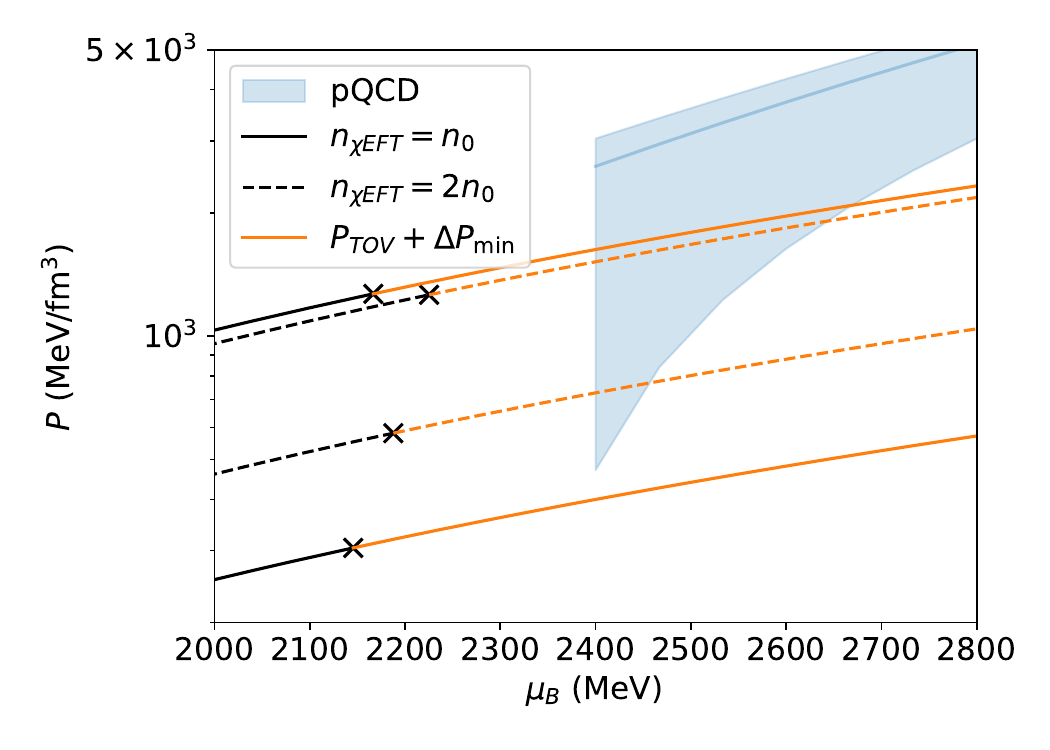}
	\caption{The $\Delta P_\mathrm{min}$ bound \cref{eq:pminbound} applied to the maximally stiff and soft NS EOSs. The blue band shows the range of pQCD predictions and is the same as in \cref{fig:pqcd}. 
	The solid (dashed) black lines show the limiting NS EOSs based on \ceft~ up to $n_\ceft=1.0 n_0$ ($2.0 n_0$) then switched to the maximally stiff and soft inner cores.
	They both predict $M_\tov\geq 2.0 M_\odot$, and are terminated at their respective TOV points marked by the black ``x''.
	From there, they are extended by $\Delta P_\mathrm{min}$ \cref{eq:dpmin} and yield the right-hand side of \cref{eq:pminbound}.
	These extensions are shown in orange.
	By the $\Delta P_\mathrm{min}$ bound \cref{eq:pminbound}, any NS EOS whose orange extension lies above the blue curves are incompatible with the underlying pQCD EOS.
	For the four cases shown, while \cref{eq:pminbound} is always satisfied for pQCD with $X\gtrsim1.5$, it is violated by those that predict very low $P_\pqcd$ ($X\simeq 1$).
	However, this may not be viewed as a constraint on NS EOSs due to the unknown strengths of non-perturbative effect at $\mu_\pqcd$ not included here.}\label{fig:dpmin}
\end{figure}

This extension by $\dpmin$ is shown in \cref{fig:dpmin} for the maximally stiff and maximally soft NS EOSs that follow the extremes of $2\sigma$ $\ceft$ up to $n_\ceft=n_0$ (solid black) and $n_\ceft=2n_0$ (dashed black).
The maximally soft NS EOSs are obtained by demanding $M_\tov=2M_\odot$, while for the maximally stiff cases I take $\Delta n_\mathrm{onset}\rightarrow\infty$ in \cref{eq:stiff} (turn off phase transitions) which yield the highest possible $M_\tov$ for given low-density EOS below $n_\ceft$.
The maximally stiff NS EOSs predict the highest $P_\tov$ and are near the bottom of  \cref{fig:dpmin}, whereas the maximally soft constructions lead to the opposite limit predicting $P_\tov\gtrsim 10^3~\MeV~\fmic$. 
All four EOSs are truncated at their respective TOV points marked by ``x'', where $\dpmin$ is added and the sum is shown in orange.
\Cref{eq:pminbound} demands the orange lines to lie below $P_\pqcd$ (shown in blue) in order that the underlying NS EOSs are consistent with chosen pQCD calculations.
For pQCD EOS with $X\gtrsim 2$, $P_\pqcd$ lie well above the orange curves $P_\tov+\Delta P_\mathrm{min}(\mu_\pqcd)$, suggesting all of the limiting NS EOSs satisfy \cref{eq:pminbound}.
However, if $X\lesssim 1.5$, soft NS EOSs are in tension with pQCD and appear to be ruled out by  \cref{eq:pminbound}.

The obvious caveat is that $P_\pqcd$ currently has considerable uncertainties especially near lower values of $X$ relevant for this constraint. Violations of \cref{eq:pminbound} only occur for pQCD EOSs with $X\lesssim 1.3$, or roughly $P_\pqcd(2.4\GeV)\lesssim 1.3$ GeV/fm$^3$. The partial N3LO contribution raises $P_\pqcd$ and renders this constraint weaker, see \cref{fig:dpmin_n3lo} in \cref{sec:n3lo}.

The central issue underlying the above argument is the assumption that the unpaired quark matter described by \pqcd~ is the true ground state around $\mu_\pqcd\simeq 2.5-3~\GeV$.
Below $\bar\Lambda\sim\mu_q\simeq \GeV$ the strong coupling constant $g_s=\sqrt{4\pi\alpha_s}$ that appears in the Lagrangian is above 1, suggesting non-perturbative effects could still dominate in this regime.
Since the determination of ground state requires knowledge of all possible phases
one is unable to make the call based on the pQCD EOS alone, even if its uncertainty can be significantly reduced.

In fact, the dense quark matter EOS is known to receive non-perturbative contributions from the pairing of quarks near their Fermi surface.
Akin to Copper pairs in electric superconductors, this gives rise to color superconductivity ~\cite{Alford:1997zt,Son:1998uk,Alford:1998mk,Schafer:1999jg,Rajagopal:2000wf,Alford:2007xm} in the cold quark matter (CQM). 
Assuming the ground state at $\mu_B\gtrsim2.5~\GeV\gg 3m_s$ is the most symmetric color-flavor-locked (CFL) superconducting phase,
the pressure and density of the CQM are given by
\begin{equation}\label{eq:cfleos}
\begin{aligned}
P_\cqm=P_\pqcd + P_\cfl,\\
n_\cqm=n_\pqcd + n_\cfl,
\end{aligned}
\end{equation}
where
\begin{equation}\label{eq:pcfl}
P_\cfl=\frac{\Delta_\cfl^2 \mu_B^2}{3\pi^2}\simeq0.3 \mathrm{GeV/fm}^3~\lrp{\frac{\Delta_\cfl}{100~\MeV}}^2\lrp{\frac{\mu_B}{2.4~\GeV}}^2,
\end{equation}
$n_\cfl=\pd P_\cfl/\pd \mu_B$, and $\Delta_\cfl$ is the pairing gap which from the leading order $\mathcal{O}(g_s)$ gap equation is given by $\Delta_\cfl/\mu_q\sim\exp(-\const/g_s)$. 
While estimates based on leading-order pQCD calculations suggest a modest gap $\Delta_\cfl\lesssim 10-50~\MeV$~\cite{Alford:2007xm}, those rooted in phenomenological models reveal much larger possibilities $\Delta_\cfl\sim50-200$ MeV at $\mu_B\simeq3$ GeV~\cite{Berges:1998rc,Carter:1998ji,Pisarski:1999bf,Rajagopal:2000wf,Alford:2007xm,Braun:2021uua}. These values can vary by factors of a few in either direction depending on the color/flavor/spin structures of Cooper pairs, higher order corrections, and if non-standard pairings emerge~\cite{Alford:2007xm}.

Taking into account the pairing contribution, the $\dpmin$ bound \cref{eq:pminbound} should really read
\begin{equation}\label{eq:dpminboundCFL}
P_\pqcd + P_\cfl \geq P_\tov+\Delta P_\mathrm{min}(\mu_\pqcd).
\end{equation}
It becomes apparent that the tension in \cref{fig:dpmin} is resolved if the deficit in $P_\pqcd$ is compensated by $P_\cfl\simeq 600~\MeV~\fmic$ at $\mu_B=2.4~\GeV$, which translates to $\Delta_\cfl\simeq 200~\MeV$.
When the partial N3LO contribution is included, the deficit in pressure reduces and only requires $\Delta_\cfl\simeq 100~\MeV$, see \cref{fig:dpmax_n3lo}.

\begin{figure}
	\includegraphics[width=0.92\linewidth]{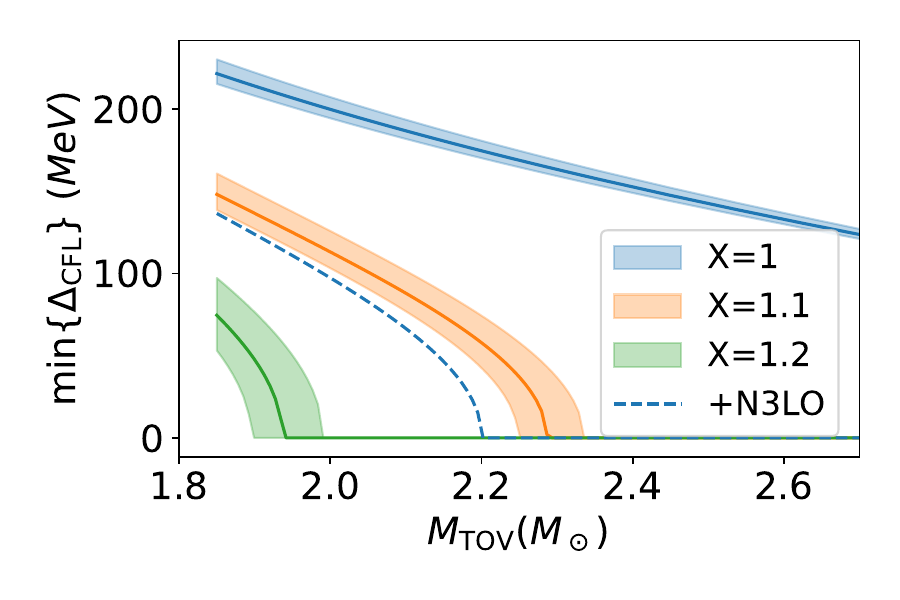}
	\caption{Lowest superconducting gaps at $\mu_B=2.4~\GeV$ required to reconcile tensions with the $\dpmin$ bound for the maximally soft NS EOSs.
	These values are robust upper limits on the RHS of \cref{eq:mingap}, assuming $\ceft$ is valid up to $2n_0$.
	That is to say, at given $M_\tov$, violations of the $\dpmin$ bound by {\em any} NS EOSs can be explained by pairing gaps no greater than the values shown here. 
	The colors represent choices of the pQCD renormalization scale $X$, and the bands correspond to the $2\sigma$ $\ceft$ uncertainties.
	The figure is cut off below $M_\tov\leq1.86M_\odot$, the $3\sigma$ lower limit of the inferred mass of PSR J0740+6620~\cite{NANOGrav:2019jur,Fonseca:2021wxt}.
	}\label{fig:mingap}
\end{figure}

The minimum pairing gap $\min\{\Delta_\cfl\}$ required to render {\em any} NS EOS that supports at least $2M_\odot$ compatible with pQCD can be found by rearranging \cref{eq:dpminboundCFL}, which leads to a lower bound on $P_\cfl$:
\begin{multline}\label{eq:mingap}
 P_\cfl(\mu_B) \geq P_\tov-P_\pqcd(\mu_B) +\Delta P_\mathrm{min}(\mu_B)\\
 \equiv\Delta P(\mu_B)+\dpmin(\mu_B).
\end{multline}
The right-hand side (RHS) of the above inequality has a global maximum across all NS inner core EOSs, and this maximum is reached by the maximally soft NS EOS~\cref{eq:soft}.
I show this upper bound in \cref{fig:mingap} as a function of the NS maximum mass.
As noted earlier, violations of $\dpmin$ may only occur if $X\lesssim1.2$, and with the partial N3LO contribution included the threshold tightens to $X\lesssim1.1$ (dashed line).
For the N2LO pQCD EOS, gaps less than $\simeq200~\MeV$ are adequate to resolve {\em any} violations of \cref{eq:pminbound} given the existence of two-solar-mass pulsars, and with the N3LO term included, $\Delta_\cfl\lesssim 100~\MeV$ would suffice.
It is for this reason I caution that $\dpmin$ shall not be considered as a bound on the NS EOS until the non-perturbative $\Delta_\cfl$ can be pinpointed exactly.

On the other hand, if higher order terms in pQCD narrow down $P_\pqcd(2.4~\GeV)$ to be below $\simeq1.3~\GeV/\text{fm}^{3}$,  astrophysical evidence for a rather soft NS inner core EOS could indicate non-perturbative phenomena around $\mu_B\simeq 2.4-3~\GeV$.
Assuming no additional repulsion is found in the cold quark matter,
violations of the $\dpmin$ bound \cref{eq:pminbound} would strongly support a superconducting gap $\Delta_\cfl\simeq50-200~\MeV$ {\em above} neutron star densities without directly probing it.

These at-best mild violations of the $\dpmin$ bound \cref{eq:pminbound} are due to the fact that $P_\tov$ is bounded from above. This upper limit is given by the maximally soft NS inner core EOSs.
The existence of $2M_\odot$ pulsars robustly rules out softer EOSs that predict higher $P_\tov$, e.g.,~\cite{Drischler:2020fvz}.
Had I not taken into account the two-solar-mass pulsars, much stronger violations of $\dpmin$ would have been allowed and consequently larger minimal superconducting gaps $\mathrm{min}\{\Delta_\mathrm{CFL}\}$ required to compensate these deficits.
Our model-independent approach utilizing \cref{eq:soft} makes this fact explicit and robust.
Furthermore, these results are insensitive to astrophysical uncertainties on $M_\tov$.
Even if $M_\tov$ is at the lower $3\sigma$ inferred mass of PSR J0740+6620 $M\approx 1.87M_\odot$~\cite{NANOGrav:2019jur,Fonseca:2021wxt}, the above discussion remains valid (see \cref{fig:mingap}).

\subsection{the $\Delta P_\mathrm{max}$ bound}\label{sec:dpmax}

\begin{figure}
	\includegraphics[width=\linewidth]{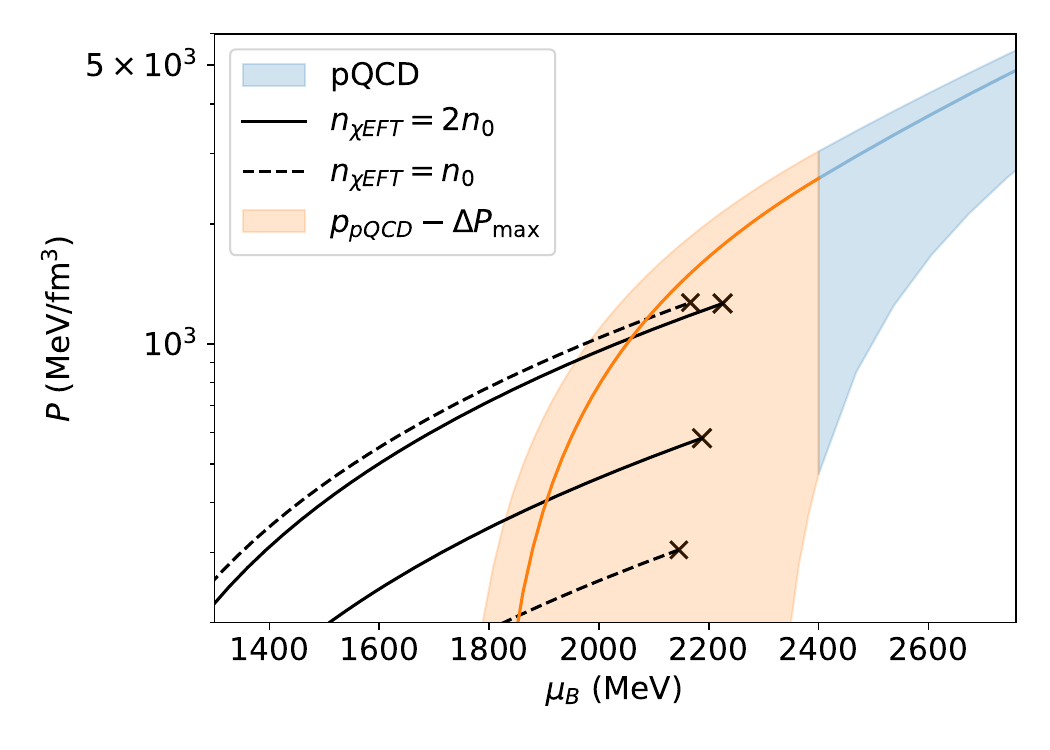}
	\caption{The $\Delta P_\mathrm{max}$ bound \cref{eq:pmaxbound} confronting the maximally stiff and maximally soft NS EOSs. The blue band shows the range of pQCD predictions up to N2LO, and is the same as in \cref{fig:pqcd}. The solid (dashed) black lines depict NS EOSs following \ceft~ up to $n_\ceft=
	n_0$ ($2 n_0$) then switched to the maximally stiff and soft inner cores. They are truncated at their respective TOV points, marked by the black ``x''.
	The RHS of \cref{eq:pmaxbound} is shown in orange and is matched to pQCD at $\mu_\pqcd=2.4$ GeV.
	The four limiting NS EOSs are only compatible with \cref{eq:pmaxbound} if $X\lesssim1.3$. They are ruled out by a considerable region of the pQCD parameter space at $\mu_\pqcd =2.4$ GeV.
	The brown stars labeled ``critical point'' mark the lowest densities at which the bound \cref{eq:pmaxbound} is violated. 
	Although they lie quite deep inside the TOV points suggesting large sections of the EOSs are ruled out, 
	these excluded segments are only concern stars very close to the TOV limit (see the red segments in \cref{fig:MR_maxss}).
	}\label{fig:dpmax}
\end{figure}

For a chosen pQCD EOS at high-density matching point, $\dpmax$ \cref{eq:dpmax} only depends on $\mu_L$.
It is therefore convenient to impose $\dpmax$ at the high-density endpoint and compare the resulting $P_\pqcd-\dpmax$ with $P_L$ at the low-density endpoint, requiring
\begin{equation}\label{eq:pmaxbound}
P_L \geq P_\pqcd-\Delta P_\mathrm{max}(\mu_L).
\end{equation}

As in the previous subsection, I first consider the limiting NS inner core EOSs stitched at either $n_\ceft=2n_0$ (solid in \cref{fig:dpmax}) or $n_\ceft=n_0$ (dashed).
They are shown in black in \cref{fig:dpmax} and are truncated at their respective TOV points.
The RHS of \cref{eq:pmaxbound} is depicted in orange.
The bound \cref{eq:pmaxbound} then rules out any NS EOS that lies above given orange curve, as it is incompatible with the assumed pQCD EOS upon which the $\dpmax$ extension under consideration is based.
It is clear from \cref{fig:dpmax} that all four limiting NS EOSs are in tension with pQCD predictions at $\mu_\pqcd=2.4$ GeV for $X\gtrsim 1.6$. 
This appears to be quite remarkable since the maximally soft EOSs predict the highest possible pressure $P_\tov$, and yet even these extreme values at around $\mu_B\simeq2.2$ GeV are still too low to reach the majority of pQCD EOSs at $\mu_B=2.4$ GeV.
On the other end of the spectrum, predictions of $P_\tov$ by the maximally stiff NS EOSs are the lowest possible and therefore in stronger tension with a wider range of pQCD calculations. 
However, owning to the untamed uncertainties associated with $P_\pqcd$, even these extremely low values of $P_\tov$ cannot be ruled out completely as they are still compatible with \cref{eq:pmaxbound} when $X\lesssim 1.3$.

\begin{figure}
	\includegraphics[width=0.92\linewidth]{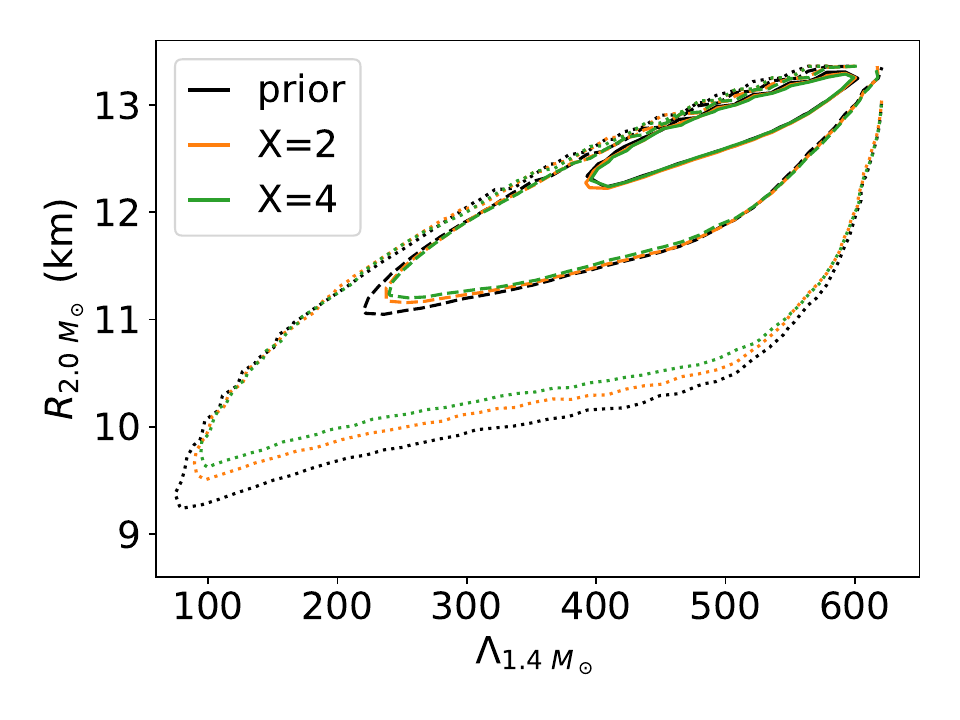}
	\caption{
	Confidence intervals (CIs) on $\Lambda_{1.4\msol}$ and $R_{2.0\msol}$ before and after imposing the $\dpmax$ bound at $\mu_\pqcd=2.4$ GeV assuming $X=2$ (orange) and $X=4$ (green).
	The solid, dashed, and dotted lines represent the $50\%$, $90\%$, and $99\%$ CIs.
	The underlying EOS samples are generated following the $\ceft$-based low-density model up to $n_\ceft=2n_0$ and are described in \cref{sec:nseos}.
	Only the existence of two-solar-mass NSs is imposed here. 
	Using the N3LO $\ceft$ EOS~\cite{Drischler:2020yad} up to $n_\ceft=2n_0$ automatically implies $\Lambda_{1.4\msol}\lesssim650$ (see also~\cite{Drischler:2020fvz}).
	The posteriors are not appreciably different from the prior, suggesting the $\dpmax$ constraint is mostly orthogonal to current or expected astrophysical bounds.
	}\label{fig:l14r20}
\end{figure}

Since the maximally stiff and maximally soft inner cores delimit the space of all physical EOSs,
it immediately follows that a wide range of NS EOSs may be at risk of being constrained by \cref{eq:pmaxbound}.
And for those that are in tension with pQCD, they cannot be ruled out completely because the maximally stiff NS EOSs are yet to be excluded robustly due to the large pQCD truncation errors manifested as a strong sensitivity towards $X$.

\Cref{fig:l14r20} shows predictions for the radius of $2M_\odot$ NSs and the tidal deformability of $1.4M_\odot$ stars for about 10 million randomly generated samples following \ceft~ up to $2n_0$.
Upon imposing \cref{eq:pmaxbound} between the EOS-dependent $\mu_\tov$ and $\mu_\pqcd=2.4$ GeV, about $20\%$ of the samples in this pool are excluded\footnote{The fraction of samples excluded by pQCD constraints vary from $\sim10\%$ to $\sim60\%$ across different subsets of NS inner core parameterizations, but the effects on observables and EOSs are almost always insensitive to these choices. }.
I take $X=2$ and $X=4$ as they lead to the highest $P_\pqcd$ and consequently the most optimistic scenario for the $\dpmax$ constraint.
This constraint neither affects appreciably the bounds on $R_{2.0\msol}$ or $\Lambda_{1.4\msol}$, nor the correlations between the two observables.
While there is a minor preference for slightly higher $R_{2.0\msol}$ in the posterior, as the excluded samples tend to cluster around the lower end of $R_{2.0\msol}$ 
(because the central baryon chemical potentials at $2M_\odot$ is further away from that at $M_\tov$ for stiffer EOSs, see below), this shift is less than $\approx0.2$ km.

\begin{figure}
	\includegraphics[width=0.92\linewidth]{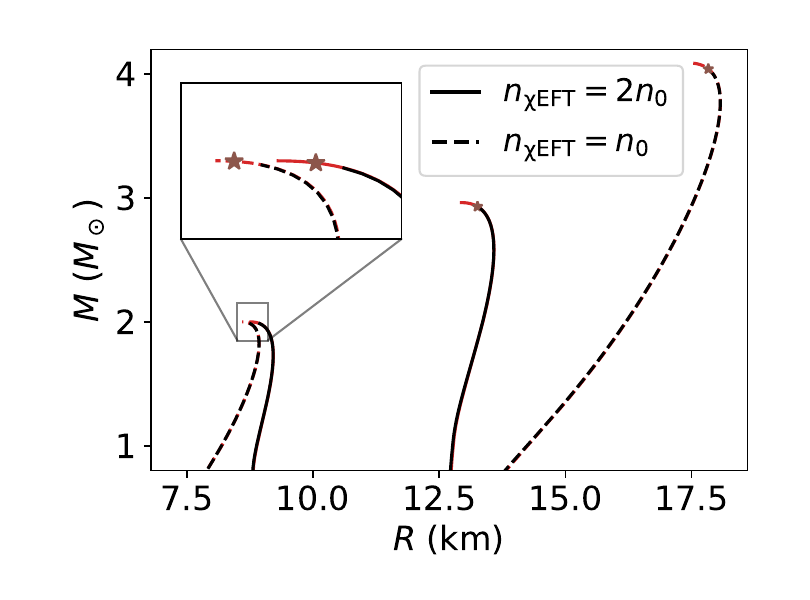}
	\caption{Mass-radius relations for the maximally stiff and maximally soft NS EOSs. Only the segments shown in red are excluded by the $\dpmax$ bound imposed at $\mu_B=2.4~\GeV$ assuming $X=4$. The rest of the curves depicted in black are compatible with the chosen $\pqcd$ EOS. The onsets of the red segments correspond to the critical points for the underlying  NS EOSs,
	and those assuming $X=2$ are marked by brown stars as in \cref{fig:dpmax}. 
	In any case, no NSs further than $\sim0.06~M_\odot$ away from $M_\tov$ are impacted by \cref{eq:pmaxbound}.
	}\label{fig:MR_maxss}
\end{figure}

To understand better the feeble impacts on NS global observables by the $\dpmax$ bound \cref{eq:pmaxbound},
let us explicitly examine the segments of NS EOSs affected by \cref{eq:pmaxbound}.
I begin by looking at the limiting scenarios shown in \cref{fig:dpmax}.
As explained earlier, violations of \cref{eq:pmaxbound} occur when an NS EOS crosses and lies above the chosen orange curve that is specific to a chosen pQCD renormalization scale $X$.
Therefore, for any NS EOS that is in tension with \cref{eq:pmaxbound}, only the segment above the crossing point is excluded. 
The low-density part preceding the crossing remains valid.  
The crossings will be referred to as the ``critical'' points.

In \cref{fig:dpmax} the critical points of the four limiting NS EOSs are marked by the brown stars assuming $X=2$.
Although significant portions of these EOSs lie above the critical point where the baryon chemical potential $\mu_\crit\simeq1.8-2.0$ GeV, these high-density segments are only probed by stars very close to the maximum mass $M_\tov$.
Excluded segments of the mass-radius curves  are shown in red in \cref{fig:MR_maxss}.
For the maximally soft NS EOSs that support $M_\tov=2.0M_\odot$, the lowest masses impacted by the aforementioned constraint are within $0.01M_\odot$ to $M_\tov$.
The maximally stiff NS EOSs are more susceptible to such constraints, yet only NSs heavier than about $2.93\msol$ ($4.04\msol$) for the one following $\ceft$ up to $n_\ceft=2n_0$ ($n_\ceft=n_0$) are affected.
Considering their predictions for the maximum mass are $M_\tov=2.96\msol$ when $n_\ceft=2n_0$ and $M_\tov=4.09\msol$ when $n_\ceft=n_0$, it appears unlikely that these minor effects may lead to detectable consequences. 
Below, for brevity, I shall refer to the lowest NS mass impacted by pQCD via \cref{eq:pmaxbound} as $M_\crit$, i.e., the mass of the star whose central density is the critical point.

The radius of the star at the critical point $R_\crit\equiv R(M_\crit)$ is also very close to $R_\tov$.
For the maximally soft NS EOS with $n_\ceft=n_0$ ($2n_0$), $R_\crit$ is about $0.1$ km ($0.1$ km) larger than $R_\tov=8.6$ km ($8.9$ km), 
and for the maximally stiff EOS following \ceft~ up to $n_0$ ($2n_0$), $R_\crit$ is about $0.4$ km ($0.3$ km) larger than $R_\tov=17.4$ km ($12.9$ km)
\footnote{Predictions of $R_\tov$ by the maximally stiff NS EOSs are not a bound.}.
This compounded with the nearly flat slopes $\ud M/\ud R$ near the TOV limit strengthens the finding that pQCD is impotent in placing bounds on NS static observables. 

\begin{figure}
	\includegraphics[width=0.92\linewidth]{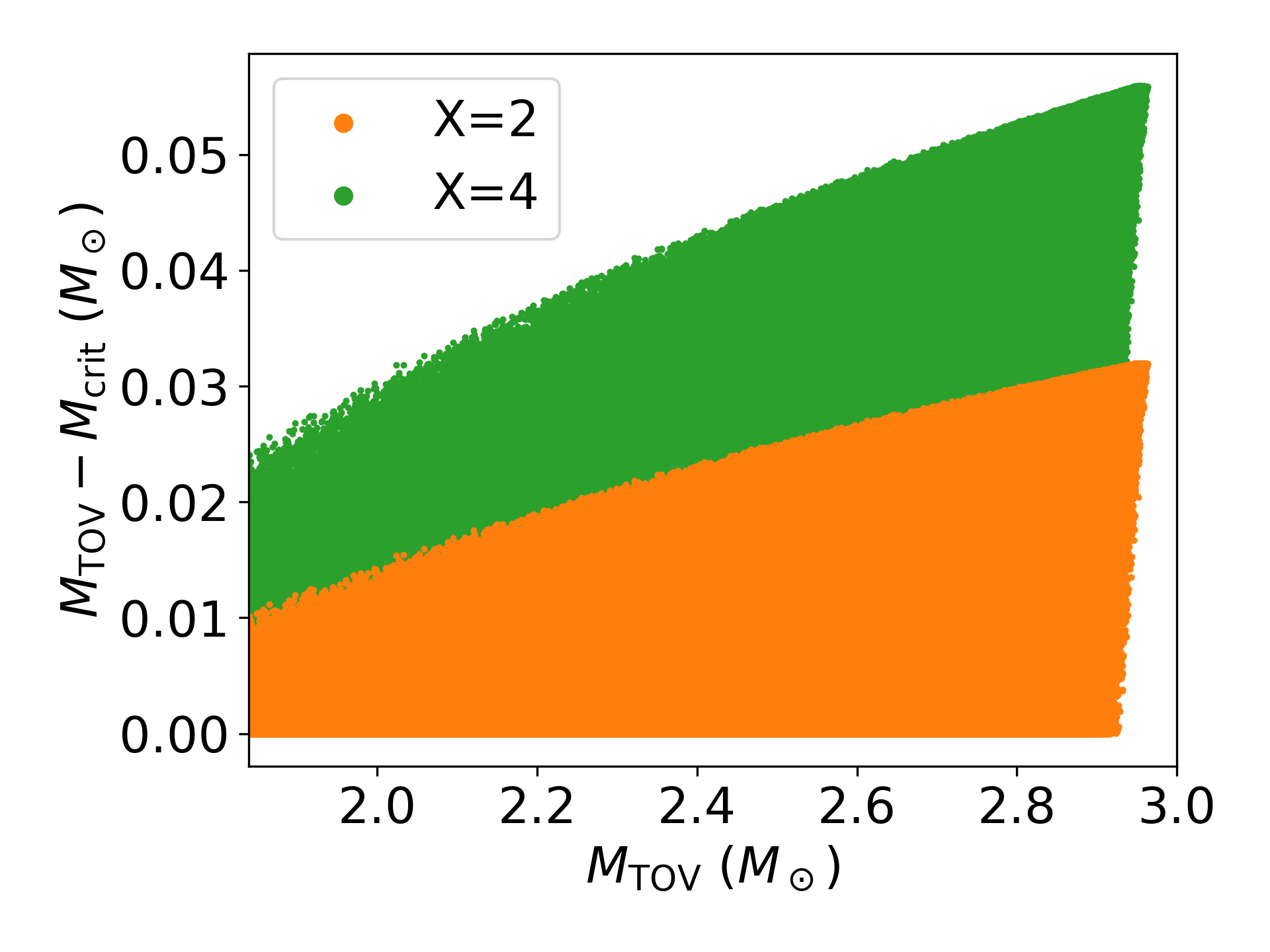}
	\caption{The distance between $M_\crit$ and $M_\tov$ for samples excluded by \cref{eq:pmaxbound} in \cref{fig:l14r20}.
	They are also the minimal changes in $M_\tov$ needed by the excluded NS EOSs to resolve their tension with \cref{eq:pmaxbound}.
	These very short distance in NS masses $M_\tov-M_\crit$ between the critical point and the TOV point, combined with their considerable separations in baryon chemical potentials $\mu_\tov-\mu_\crit$ (see \cref{fig:dpmax,fig:maxCs}), suggest that sufficiently flexible parameterizations of the NS EOS would not observe noticeable impacts on the size of NSs, since the EOS at TOV points (which \cref{eq:pmaxbound} directly constrains) would largely decouple from that at much lower $\mu_B$ (which astrophysics probes).
	}\label{fig:Mcrit}
\end{figure}

Since all possibilities in the mass-radius plane are bracketed by predictions of the limiting NS EOSs~\footnote{except for dwarf neutron stars \cite{Zhou:2023cvq}.},
it is safe to conclude in the absence of additional assumptions that \cref{eq:pmaxbound} only impacts stars in the vicinity of $M_\tov$.
\Cref{fig:Mcrit} shows the proximity $M_\tov-M_\crit$ of critical masses to the TOV limits for excluded samples in \cref{fig:l14r20}.
For NS EOSs supporting higher $M_\tov$, pQCD constraints radiate further downwards but the effect is confined to a very limited range of NS masses throughout.
Indeed, this is a ubiquitous feature shared by every solutions to the TOV equation.
For all the EOS samples considered, the derivatives of NS masses with respect to their central baryon chemical potential are small $\left(\ud M/\ud \mu_c\right)_{2\msol}\lesssim 10^{-2}~\msol/\MeV$ for two-solar-mass stars and decreases rapidly at higher NS masses.\\

Our discussion above on the critical points hints at ways of modifying an excluded NS EOS to reconcile it with high-density inputs.
Since the segment below $\mu_\crit$ already complies with \cref{eq:pmaxbound}, the minimal modification would be to follow the original EOS up to the critical point, then stick to the $\dpmax$ extension $P(\mu_B)=P_\pqcd-\Delta P_\mathrm{max}(\mu_B)$ to avoid dropping below it.
Implementations of this procedure entail a first order phase transition at $\mu_\crit$ (which gives rise to the sharp corner that bends the $P(\mu)$ curve upwards), followed by a section of constant $C_s=\cmax=1$.
Notice that this is exactly the scenario depicted in red in \cref{fig:construction}, and the number density immediately after the phase transition is given by \cref{eq:nLmax}.
A simple estimate noting that $n_\pqcd\simeq30-50 n_0$ and $\mu_\crit/\mu_\pqcd\gtrsim 0.5$ suggests the required jump in densities associated with the first order phase transition is typically huge ($\Delta n_B\gtrsim10 n_0$, $\Delta\mathcal{E}\gtrsim$ GeV/$\mathrm{fm}^3$, see \cref{fig:nconstraint} in \cref{sec:intcnstr}), rendering stable NSs beyond the critical points infeasible.
The outcome is an updated TOV limit that sits at $M_\tov^\mathrm{new}=M_\crit$.

I perform this modification for the excluded samples in \cref{fig:l14r20} and examine their updated predictions for astrophysical observables.
As discussed earlier and shown in \cref{fig:Mcrit},  changes in the maximum mass $\Delta M_\tov=M_\tov-M_\crit$ and shifts in the radius of the maximum-mass stars $\Delta R_\tov=R_\tov-R_\crit$ are moderate at best.
The relative differences $\Delta M_\tov/M_\tov$ and $\Delta R_\tov/R_\tov$ are both less than $\approx5\%$ for all the samples.
Furthermore, neither the central value nor the probability distributions of $R_\tov$ are appreciably affected, although those for $M_\tov$ are shifted downwards moderately with $\Delta M_\tov\lesssim 0.3 M_\odot$. This latter effect is prominent for samples predicting $M_\tov\gtrsim 3M_\odot$ and can be somewhat sensitive to the priors. But in all scenarios the shift would not exceed $\approx0.5M_\odot$.
Therefore, even if the two-solar-mass pulsars observed so far turn out to be close to the actual TOV limit,
pQCD considerations are still unable to affect the inference of their global properties appreciably. 
This is further confirmed by the probability distributions of both $R_\tov$ and $R_{2M_\odot}$ while demanding $M_\tov<2.3 \msol$~\cite{Demorest:2010bx,Antoniadis:2013pzd,NANOGrav:2019jur,Romani:2021xmb}, in the absence of this modification. For details see \cref{sec:astro}.\\

\begin{figure}
	\includegraphics[width=0.92\linewidth]{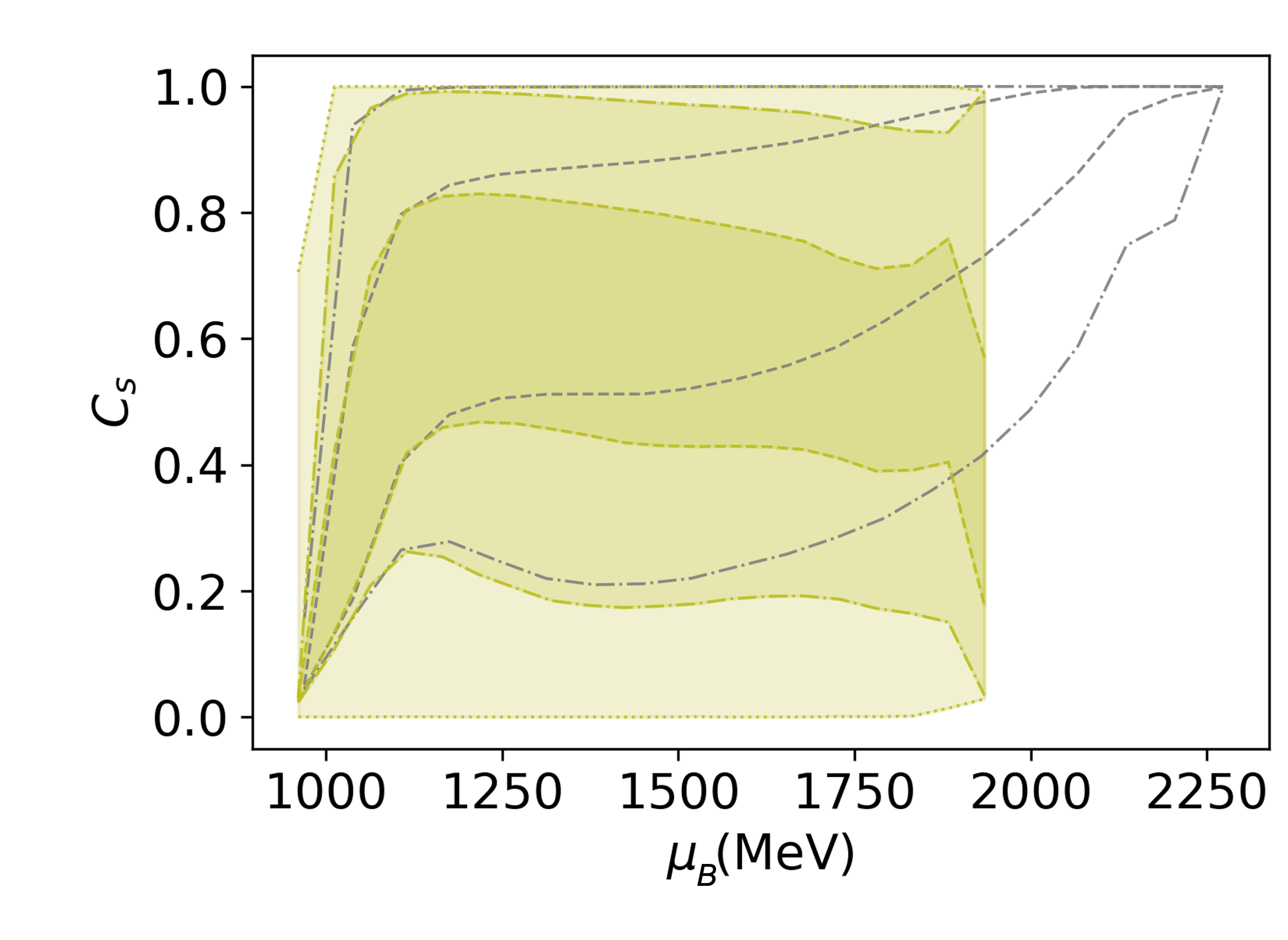}
	\caption{The prior and posterior of the speed of sound squared $C_s$ for the pQCD constraint in \cref{fig:l14r20}. The yellow dashed, dash-dotted, and dotted lines are the $50\%$, $90\%$, and $100\%$ posterior CIs respectively, whereas the gray lines show the $50\%$ and $90\%$ priors.
	The prior appears to be biased toward larger $C_s$, but this is a consequence of truncating each EOS at its TOV point, and only EOSs with larger $C_s$ predict higher $\mu_\tov$ (see \cref{fig:maxCs}).
	That the prior is indeed flat is confirmed in \cref{fig:CsN}.
	This shift in the posterior CIs shown here appears to be the strongest effect \cref{eq:pmaxbound} could impart to the inference of NS EOSs.
	}\label{fig:muCs}
\end{figure}

The modifications described above suggest that \cref{eq:pmaxbound} favors low sound speed near the TOV point. 
This preference is shown in \cref{fig:muCs}, where posteriors on the squared sound speed $C_s$ are shifted downwards by the aforementioned pQCD constraint, an increasingly noticeable effect towards higher baryon chemical potentials.
In \cref{sec:astro} I show that this shift is almost orthogonal to constraints expected from astrophysical observations.

It is striking that the pQCD constraint completely rules out $\mu_\tov\gtrsim1.9$ GeV in the posterior in \cref{fig:muCs}.
But this is not surprising considering that even the maximally soft NS EOSs,
whose predictions for $P_\tov$ are the highest, 
are excluded in \cref{fig:dpmax} if $X\gtrsim1.6$.
In other words,
the bound \cref{eq:pmaxbound} demonstrates higher sensitivity toward $\mu_\tov$ over $P_\tov$, in the sense that an NS EOS is more likely in tension with pQCD if it accommodates higher $\mu_\tov$, almost independent of its prediction for $P_\tov$, as compared with those predicting lower $\mu_\tov$.
For example, when matched to pQCD at $\mu_\pqcd=2.4$ GeV, all of the NS EOSs excluded by \cref{eq:pmaxbound} predict $\mu_\tov\gtrsim 1.8$ GeV.
These observations help establish a simple criterion on the applicability of \cref{eq:pmaxbound}.
I find that the maximum speed of sound squared $\cmax$ of an NS EOS informs $\mu_\tov$.
Assuming $\ceft$ is valid up to $n_\ceft=n_0$ (2$n_0$), $\cmax$ needs to exceed $\approx$ 0.5 (0.6) in order that $\mu_\tov\gtrsim1.8$ GeV.
This correlation also explains the seemingly biased sample selections shown in \cref{fig:muCs}: 
the prior on $C_s$ favors higher $C_s$ with increasing $\mu_B$ because only EOSs with larger $\cmax$ would lead to higher $\mu_\tov$, and if the EOSs are not truncated at their respective TOV points the prior would be mostly flat and featureless.
The CIs plotted against number densities shown in \cref{fig:CsN} also corroborate this.

\begin{figure}
	\includegraphics[width=0.92\linewidth]{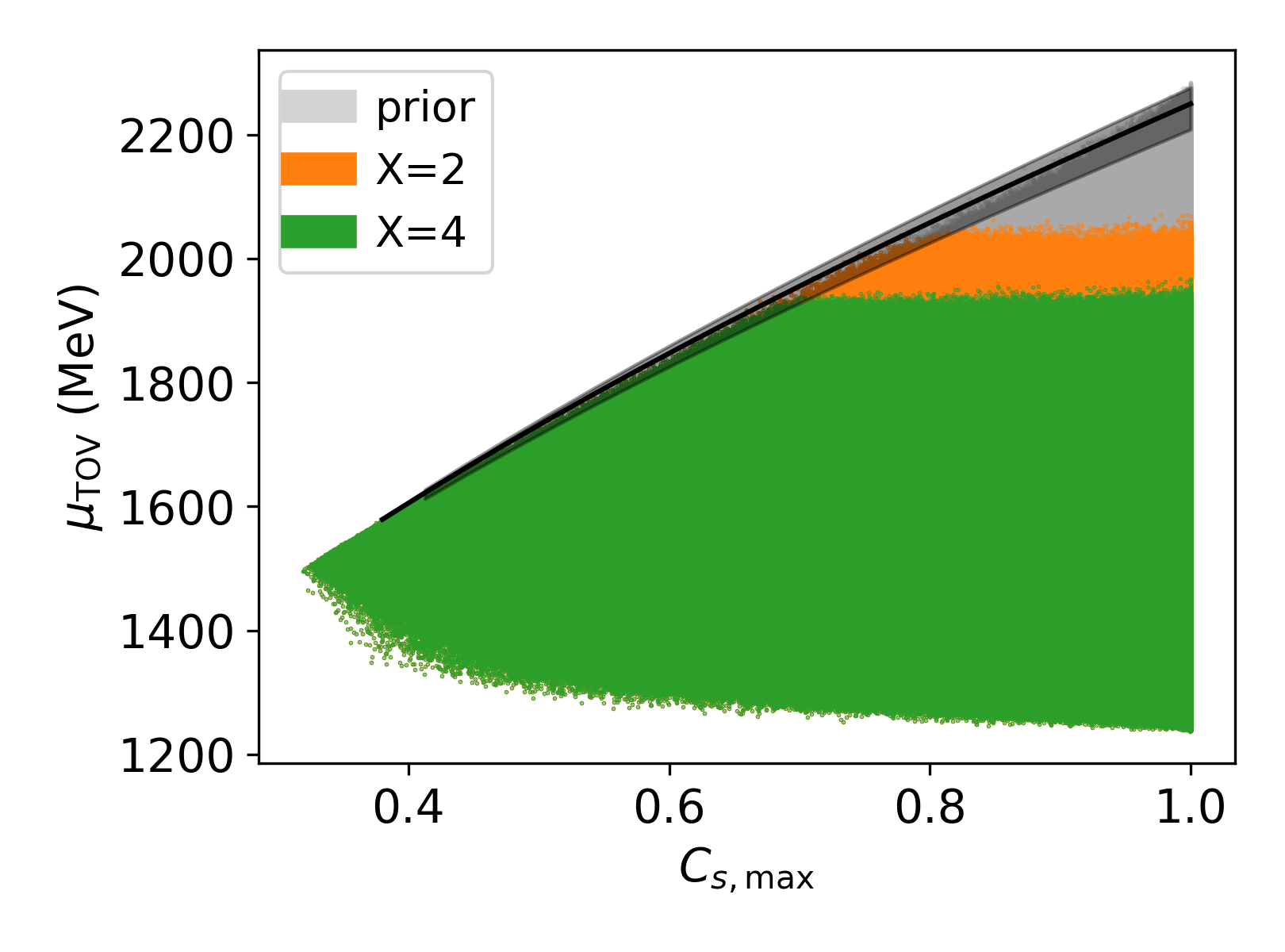}
	\caption{$\cmax\gtrsim0.6$ inside NSs is a necessary condition for NS EOSs constrained by the $\dpmax$ consideration.
	The collection of EOS samples is shown in gray.
	The black curve is based on the maximally soft NS EOSs and yield the highest possible $\mu_\tov$ at fixed $\cmax$. Its surrounding bands reflect the $\pm2\sigma$ $\ceft$ uncertainties.
	Upon imposing the $\dpmax$ bound at $\mu_\pqcd=2.4~\GeV$ the posteriors for $X=2$ and $X=4$ are shown in orange and green respectively.
	All of the excluded samples predict $\mu_\tov\gtrsim1.8~\GeV$, requiring $\cmax\gtrsim0.6$ inside  NSs.}\label{fig:maxCs}
\end{figure}

I show the correlation between $\cmax$ and $\mu_\tov$ in \cref{fig:maxCs}.
For the strongest constraint with $X=4$ (green), \cref{eq:pmaxbound} becomes relevant starting at $\mu_\tov\gtrsim1.8~\GeV$ (see also \cref{fig:dpmax}), and rules out all NS EOSs with $\mu_\tov\gtrsim1.9~\GeV$.
For given $\cmax$,  the maximally soft NS EOSs shown as the black curves give the upper bound on $\mu_\tov$ where the bands correspond to the $2\sigma$ $\ceft$ uncertainties.
They reach $\mu_\tov=1.8~\GeV$ at around $\cmax=0.6$ thus set a lower bound on $\cmax$ required for any EOS to be ``close enough'' to pQCD densities.
The upshot is that an NS EOS cannot violate \cref{eq:pmaxbound} unless $\cmax\gtrsim 0.6$ inside NSs.
This is only a necessary condition, and $\cmax\gtrsim 0.6$ does not automatically lead to the violation of \cref{eq:pmaxbound}.\\

Up until now, I have focused on the most optimistic scenarios where the matching to pQCD is performed at $\mu_\pqcd=2.4$ GeV, and have chosen $X=4$ which yields the highest pressure $P_\pqcd$. The constraints weaken drastically if the matching point is raised to higher baryon chemical potentials. 
For \cref{eq:pmaxbound} imposed at $\mu_\pqcd=2.6$ GeV, only NS EOSs with $\cmax\gtrsim0.75$ can potentially be ruled out, a condition tightens to $\cmax\gtrsim 0.9$ for matching at $\mu_\pqcd=2.8$ GeV. 
If pQCD is imposed above $\mu_B\approx2.9$ GeV, \cref{eq:pmaxbound} becomes completely irrelevant.
Furthermore, lowering the renormalization scale $X$ also reduces the strength of such constraints.
At $X=1$, no NS EOSs that support $2\msol$ can be ruled out, even with soft N3LO contribution to the dense quark matter EOS included.

To account for the pQCD truncation error manifested as dependencies on $X$, 
I perform a simple inference by noting that while uncertainties associated with $P_\pqcd$ are huge, those of the number densities are controlled. 
I therefore fix $n_\pqcd$ to the fiducial value $n_B=5~\fmic$, but sample $P_\pqcd$ randomly from a uniform (and log-uniform) distribution on $[0.5, 3]$ GeV/fm$^{3}$, the approximate range of pQCD predictions at $\mu_\pqcd=2.4$ GeV with $X\in[1,4]$.
Analysis in \cref{fig:muCs} is then repeated but with this mocked $P_\pqcd$ data instead.
Unsurprisingly, the posteriors on $C_s$ are almost identical to the priors even for $\mu_B\gtrsim2.0~\GeV$, a result independent of values of $\mu_\pqcd$, whether varying $n_\pqcd$, or if the N3LO contributions are included (\cref{sec:n3lo})
\footnote{Priors favoring large $P_\pqcd$ would help, though it is not clear if such choices are justified. I impose priors on $P_\pqcd$ as it is directly involved in the $\dpmax$ bound. Choosing instead priors on $X$ is also valid but obscures this connection, and the resulting shape of the distribution on $P_\pqcd$ might not be invariant once higher order terms in pQCD are known.}.
Since the constraints on $C_s(\mu_B)$ appear to be the strongest, and incorporating uncertainties associated with the running of $\alpha_s$ will further broaden error bands (see~\cref{sec:alphas}), 
it appears that meaningful and robust constraints on the NS EOS from pQCD is infeasible at the moment.

I conclude with cautious optimism by providing a forecast on the presumption that $P_\pqcd$ at $\mu_B=2.4$ GeV will be reliably and accurately determined.
For $P_\pqcd(2.4~\GeV)=3$ GeV/fm$^3$, the median value pf the squared sound speed at the center of maximum-mass stars, $C_{s,\tov}$, is expected to be lowered by up to 0.2-0.3 by \cref{eq:pmaxbound}, whereas for $P_\pqcd(2.4~\GeV)=2$ GeV/fm$^3$ the anticipated effect drops to about 0.1-0.2, and would be less than $\sim0.05$ if $P_\pqcd(2.4~\GeV)=1.5$ GeV/fm$^3$ hardly distinguishable from systematic uncertainties due to assumptions in the underlying NS EOSs.
Again, these forecasted bounds are almost orthogonal to expected astrophysical constraints, unless highly specific features are assumed for NS EOSs.

\subsection{\cref{eq:pmaxbound} is the necessary and sufficient condition} 

As noted above, compliance with \cref{eq:pmaxbound} at $\mu_L=\mu_\tov$ is the sufficient condition for a given NS EOS, and ensuring it holds for all $\mu_L<\mu_\tov$ is the necessary condition.
Hence from a practical point of view, one only needs to check \cref{eq:pmaxbound} at the TOV point of each NS EOS to determine if the constraint applies.
Discussions on the critical point $\mu_\crit<\mu_\tov$ only serve to clarify the constraining power of pQCD, and are not necessary in general.

Recently, a so-called integral constraint is proposed in ref~\cite{Komoltsev:2021jzg}. 
It is formulated in the $\mu_B-n_B$ plane and is a weaker version of the necessary condition discussed above, see~\cref{sec:intcnstr}.
Note that the way it has been applied is only approximate as the (comparatively controlled and less important for the current purpose) uncertainties associated with \ceft~ are ignored.
Even though this constraint on number density does not reveal additional information beyond what is encoded in \cref{eq:pmaxbound}, it helps visualize the phase transitions required to reconcile tensions with the $\dpmax$ bound, and enables constraints on the pressure-energy density relation~\cite{Komoltsev:2021jzg}.

\section{discussion and conclusion}\label{sec:concl}

Aided by the model-independent bounds on the EOS~\cite{Rhoades:1974fn,Koranda:1996jm,Lattimer:2000nx,Drischler:2020fvz,Drischler:2021bup}, I have shown that ab-initio cold quark matter calculations in the perturbative regime have the potential of placing robust bounds on the EOS if NSs probe baryon chemical potentials greater than about $\mu_B\sim1.8-2.0$ GeV. 
The higher the baryon chemical potential realized, the tighter the constraint becomes. 
If the pressure of NS EOS is too high, the absence of a valid construction between $\mu_\tov$ and $\mu_\pqcd$ may indicate the presence of non-perturbative effects near $\mu_B\lesssim3$ GeV.
Until the pairing gap can be reliably and accurately calculated, such consideration may not be viewed as a robust constraint on the NS EOS.

On the other hand, an NS EOS may be incompatible with high-density inputs if its prediction for the pressure is too low to reach $P_\pqcd$.
This is embodied in the $\Delta P_{\max}$ bound \cref{eq:pmaxbound}, and is the sole requirement an NS EOS needs to comply with currently.
Constraints of this type only concern the high-density tails of NS EOSs relevant for stars very close to their respective TOV limits ($M_\tov-M_\crit$ no more than $0.06 M_\odot$ for $M_\tov=4M_\odot$, a bound decreases with decreasing $M_\tov$), therefore cannot directly affect observables for most NSs.
The strongest constraint possible (at $\mu_\pqcd=2.4$ GeV, assuming $X=4$ and including the N3LO soft contributions that push $P_\pqcd$ higher) can at best rule out $R_\tov\lesssim 9.5$ km, or compactness of the maximum-mass stars $C_\tov=M_\tov/R_\tov\gtrsim 0.33$, assuming \ceft~ is valid up to $n_0$.
Even if the two-solar-mass pulsars observed to date are in fact the TOV limit, such constraints would only raise the median value of $R_{2\msol}$ by no more than $0.2$ km.
Accounting for the uncertainties of pQCD requires choosing a prior on the renormalization scale $X$.
With natural choices of priors no meaningful bounds can be derived at the moment owning to the wild uncertainties associated with $P_\pqcd$.

Although the bound \cref{eq:pmaxbound} directly applies to the pressure, it is more sensitive to predictions of $\mu_\tov$ by NS EOSs, as even the highest $P_\tov$ by the maximally soft NS EOS  at $\mu_\tov\gtrsim2$ GeV  is too low to reach $P_\pqcd$ for $X\gtrsim2$ (see \cref{fig:dpmax}).
Any NS EOS predicting $\mu_\tov\gtrsim1.8$ GeV is at risk of being ruled out by such considerations.
The necessary condition for supporting high values of $\mu_\tov$ is large $\cmax\gtrsim0.6$ inside NSs, see \cref{fig:maxCs}.

Unlike the scenarios in \cref{sec:dpmin}, the $\dpmax$ bound in \cref{sec:dpmax} is  insensitive to non-perturbative effects. This is because \cref{eq:pmaxbound} only obtains for high pressure in the quark matter $P_\pqcd\gtrsim 2$ GeV/fm$^3$ where contributions from a typical pairing gap $\Delta\simeq200$ MeV $P_\cfl/P_\pqcd\sim (\Delta_\cfl/\mu_q)^2$ would be moderate at best, whereas in \cref{sec:dpmin} only pQCD calculations predicting low pressure are relevant. 
The $\dpmax$ constraint also appears to be insensitive to the strange quark mass $m_s$. In the decoupling limit $m_s\rightarrow\infty$, the 2-flavor quark matter EOS only differs by about $~20\%$ at $\mu_B= 2.6$ GeV, moving the bands in \cref{fig:muCs} by no more than $\approx10\%$.

\begin{figure}
	\includegraphics[width=0.92\linewidth]{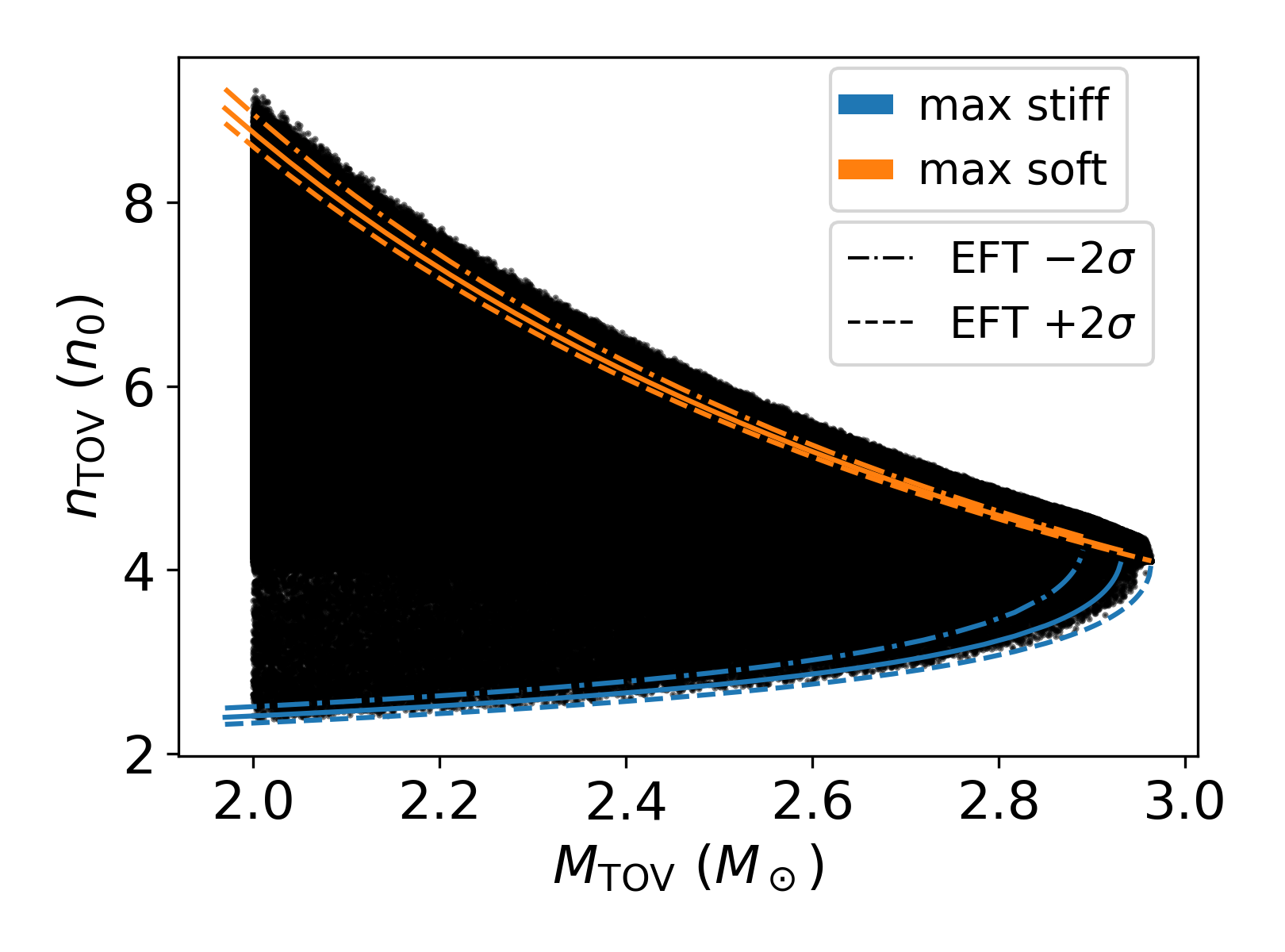}
	\caption{Correlations between $n_\tov$ and $M_\tov$. 
	Stiff NS EOSs supporting high $M_\tov$ tend to probe lower densities than those reached by their soft counterparts. 
	The lower bounds are given by the maximally stiff NS EOSs shown in blue.
	Assuming  twin stars are absent, the maximally soft NS EOS shown in orange roughly delimit the upper boundary.
	The dashed lines indicate the $2\sigma$ $\ceft$ uncertainties.
	}\label{fig:nTOV}
\end{figure}

Recently, ref \cite{Komoltsev:2021jzg} reported another {\em necessary} condition.
As discussed above and detailed in \cref{sec:intcnstr}, 
this necessary condition does not lead to additional constraints beyond \cref{eq:pmaxbound}.
By matching NS EOSs at a fixed density $n_B\simeq5-10 n_0$, that letter and the following work~\cite{Gorda:2022jvk} excluded a large number of EOSs based on their segments that are not probed by stable NSs.
This is especially problematic for stiff NS EOSs as their predictions for $n_\tov$ are typically low but for $P(n_B\simeq5-10n_0>n_\tov)$ are high.
To illustrate this point the correlation between $M_\tov$ and $n_\tov$ is shown in \cref{fig:nTOV}.
The boundaries in this figure can be reproduced by the maximally stiff (blue) and maximally soft NS EOSs (orange).
Failing to account for these correlations would unfairly bias against large values of $M_\tov$.
Ref ~\cite{Somasundaram:2022ztm} correctly addressed this issue, but arrived at the misleading conclusion that pQCD is not constraining if imposed on top of current astrophysical observations.
It has been clarified that \cref{eq:pmaxbound} is only relevant for stars very close to the TOV limit so would not affect neutron star observables appreciably, but could yield constraints on the EOS that are mostly orthogonal to astrophysical bounds.
Additionally, their claim regarding the possible interplay between pQCD constraints and $R_{2.0M_\odot}\gtrsim 13$ km is likely due to implicit biases in the parameterization of inner core EOSs, which is an approximation to a subset of the ensembles employed in this work.
For flexible parameterizations the EOS at the TOV point would largely decouple from that at lower densities, see \cref{fig:dpmax,fig:Mcrit,fig:MR_maxss}.
Addressing implicit assumptions in the parameterizations of NS EOSs is important and is relevant for a wider range of issues but requires a much much lengthier discussion and will be reported in separate work~\cite{prep1}.
Finally, we reiterate that \cref{eq:pminbound} may not be viewed as a constraint on NSs yet as it is sensitive to the undetermined strength of the superconducting gap. 
Given the existence of two-solar-mass pulsars, moderate pairing gaps $\Delta_\cfl\lesssim200~\MeV$ consistent with current estimates are sufficient to explain any possible violations.

Reducing the uncertainties of the cold quark matter EOS is crucial in materializing the potentials demonstrated in \cref{sec:dpmax}. As noted earlier, although typically referred to as the renormalization scale uncertainties, it is a form of truncation error in disguise, as the dependency on $X$ is an artifact that is expected to receive cancellations when higher order terms are known.
This is shown to be the case in dense QED~\cite{Gorda:2022zyc} and hot QCD calculations~\cite{Braaten:1995jr,Braaten:1995cm}.
One would expect it to hold in dense QCD as well assuming pQCD is a valid description down to $\mu_q\simeq\GeV$. The recently obtained  leading and next-to-leading soft contributions resummed to all orders appear to support the convergence~\cite{Fernandez:2021jfr}.
Effort employing strategies developed in \ceft~\cite{Melendez:2019izc} to understand better pQCD truncation errors is underway and will be reported in future work.

This work largely confirms previous findings that the knowledge of neutron star global properties alone may not be adequate to distinguish the relevant microscopic degrees of freedom inside the cores of NSs~\cite{Alford:2004pf}.
Although pQCD EOSs predicting high pressure can place upper bounds on the speed of sound above $\mu_B\gtrsim 2$ GeV, 
I caution against interpreting this as evidence for quark matter, as extrapolating $C_s$ from pQCD predictions above $\mu_B\sim 2.5~\GeV$ to the strongly-interacting phase at such low densities may be unwarranted.
Understanding the phase of matter in this intermediate density range is inherently challenging as neither pQCD nor NSs directly probe this regime.
The ensuing letters in this series aim to fill this gap and report model-independent studies on the ultra-dense matter.\\

Very recently, ref~\cite{Kurkela:2024xfh} performed a
 Bayesian analysis based on parameterizing the NS EOS to infer upper bounds on the superconducting gap.
I wish to point out that this constraint can be obtained model-independently with the help of maximally soft NS EOSs, see \cref{sec:maxgap}.

\section*{Acknowledgment}
I would like to thank Sanjay Reddy for discussions.
During the conception and completion of this work the author is supported 
by Grant No. PHY-1430152 (JINA Center for the Evolution of the Elements) and the Institute for Nuclear Theory Grant No. DE-FG02-00ER41132 from the Department of Energy, and
by NSF PFC 2020275 (Network for Neutrinos, Nuclear Astrophysics, and Symmetries (N3AS)).

\appendix

\section{the running of $\alpha_s$}\label{sec:alphas}

This appendix is a brief overview of the uncertainties associated with $\alpha_s\equiv g_s^2/(4\pi)^2$ around GeV scales.
To obtain the strong coupling constant at a given scale $\bar\Lambda$ I solve the renormalization group equation
\begin{equation}\label{eq:qcd_betafunc}
\diff{\alpha_s}{\log\bar{\Lambda}^2}
=-\sum_{i\geq0} \frac{\beta_{i}}{(4 \pi )^{i+1}}\alpha_s^{i+2}
=-\sum_{i\geq0} b_{i}\alpha_s^{i+2}
\end{equation}

up to 2, 3, 4 and 5 loops. For $N_c=3$ QCD with $N_f$ active flavors of quarks the beta function coefficients are given by~\cite{Baikov:2016tgj}

\begin{widetext}
\begin{align}
\beta_0&=11-\frac{2 N_f}{3},\\
\beta_1&=102-\frac{38 N_f}{3},\\
\beta_2&=\frac{325 N_f^2}{54}-\frac{5033 N_f}{18}+\frac{2857}{2},\\
\beta_3&=\frac{149753}{6}+3564\zeta_3 - \left(\frac{1078361}{162}+\frac{6508}{27}\zeta_3\right)N_f
					+\left(\frac{50065}{162}+\frac{6472}{81}\zeta_3\right)N_f^2+\frac{1093}{729}N_f^3,\\
\begin{split}
\beta_4&=\frac{8157455}{16}+\frac{621885}{2}\zeta_3-\frac{88209}{2}\zeta_4-288090\zeta_5
			+\left(-\frac{336460813}{1944}-\frac{4811164}{81}\zeta_3+\frac{33935}{6}\zeta_4+\frac{1358995}{27}\zeta_5\right)N_f\\
			{}&+\left(\frac{25960913}{1944}+\frac{98531}{81}\zeta_3-\frac{10526}{9}\zeta_4-\frac{381760}{81}\zeta_5\right)N_f^2\\
			{}&+\left(-\frac{630559}{5832}-\frac{48722}{243}\zeta_3+\frac{1618}{27}\zeta_4+\frac{460}{9}\zeta_5\right)N_f^3	
			+\left(\frac{1205}{2916}-\frac{152}{81}\zeta_3\right)N_f^4,
\end{split}
\end{align}
\end{widetext}
where $\zeta_n\equiv\zeta(n)$ are values of the Riemann zeta function.
Coefficients beyond 2 loop ($\beta_1$) are renormalization scheme dependent, and the values quoted above are given in $\msbar$. 
\Cref{eq:qcd_betafunc} can be solved either numerically, or analytically albeit approximately in a perturbative fashion.
Defining $t=2\log(\bar{\Lambda}/\bar{\Lambda}_{\msbar})$ and $l=\log(t)$, an iterative solution to the 5-loop renormalization group equation ($\mathcal{O}(\alpha_s^6)$ on the right hand side of \cref{eq:qcd_betafunc}) is given by

\begin{widetext}
\begin{multline}\label{eq:alpha5loop}
\alpha_s(\bar{\Lambda}^2)=\frac{1}{b_0 t}\left[
1-\frac{b_1 l}{b_0^2 t}+\frac{b_1^2 \left(l^2-l-1\right)+b_0 b_2}{b_0^4 t^2}+\frac{b_1^3 \left(-2 l^3+5 l^2+4 l-1\right)-6 b_0 b_1 b_2 l+b_0^2 b_3}{2 b_0^6 t^3} \right.\\
\left.
+\frac{18 b_0 b_2 b_1^2 \left(2 l^2-l-1\right)+b_1^4 \left(6 l^4-26 l^3-9 l^2+24 l+7\right)+2 b_0^2 \left(5 b_2^2+b_0 b_4\right)-b_0^2 b_1 b_3 (12 l+1)}{6 b_0^8 t^4}\right].
\end{multline}
Its exact derivatives up to the second order are
\begin{multline}\label{eq:dalpha5loop}
\diff{\alpha_s (\bar{\Lambda}^2)}{\log\bar{\Lambda}^2}=-\frac{1}{b_0 t^2}\left[
1+\frac{b_1(1-2 l)}{b_0^2 t}+
\frac{b_1^2 (-2 - 5 l + 3 l^2)+3 b_0 b_2}{b_0^4 t^2}\right.\\
+\frac{b_1^3 (-4 + 3 l + 13 l^2 - 4 l^3)+3 b_0 b_2 b_1 (1-4 l)+2 b_0^2 b_3}{b_0^6 t^3}\\
\left.
+\frac{b_1^4 (11 + 138 l + 33 l^2 - 154 l^3 + 30 l^4)+18 b_0 b_2 b_1^2 (-4 - 9 l + 10 l^2)+b_0^2 b_3 b_1 (7-60 l)+10 b_0^2 \left(5 b_2^2+b_0 b_4\right)}{6 b_0^8 t^4}\right],
\end{multline}

\begin{multline}\label{eq:ddalpha5loop}
\frac{\ud^2\alpha_s (\bar{\Lambda}^2)}{\ud\left(\log\bar{\Lambda }^2\right)^2}=\frac{2}{b_0 t^3}\left[
1+\frac{b_1 (5-6 l)}{2 b_0^2 t}+
\frac{b_1^2 (-3 - 26 l + 12 l^2)+12 b_0 b_2}{2 b_0^4 t^2}\right.\\
+\frac{-b_1^3 (23 + 11 l - 77 l^2 + 20 l^3)+3 b_0 b_2 b_1 (9-20 l)+10 b_0^2 b_3}{2 b_0^6 t^3}\\
\left.
+\frac{3 b_0 b_2 b_1^2 \left(60 l^2-74 l-15\right)+b_1^4 (127 l + 110 l^2 - 174 l^3 + 30 l^4)+b_0^2 b_3 b_1 (17-60 l)+10 b_0^2 \left(5 b_2^2+b_0 b_4\right)}{2 b_0^8 t^4}\right].
\end{multline}

\end{widetext}
I do not use the renormalization group equation \cref{eq:qcd_betafunc} to compute the derivatives 
as \cref{eq:alpha5loop} is only an approximate solution to the 5-loop renormalization group equation. The residue, although small, would spoil the thermodynamic consistency of the pQCD EOS if \cref{eq:qcd_betafunc} were to used in calculating  $n_B=\diff{P}{\mu_B}$ and $C_s=\ud\log \mu_B/\ud\log n_B$.

The 2-, 3-, 4-loop solutions of $\alpha_s$ can be obtained by setting $b_{i>n_\mathrm{loop}-1}=0$ where $n_\mathrm{loop}=2,3,4$. It is also customary, although not necessary, to truncate the iterative solutions in \cref{eq:alpha5loop} at order $\lrp{\frac{b_i}{b_0^2}}^{n_\mathrm{loop}-1}$ in the literature, i.e., only keep the leading power in $b_{n_\mathrm{loop}-1}$. This convention is adopted throughout the manuscript. 

\begin{figure}
	\includegraphics[width=0.98\linewidth]{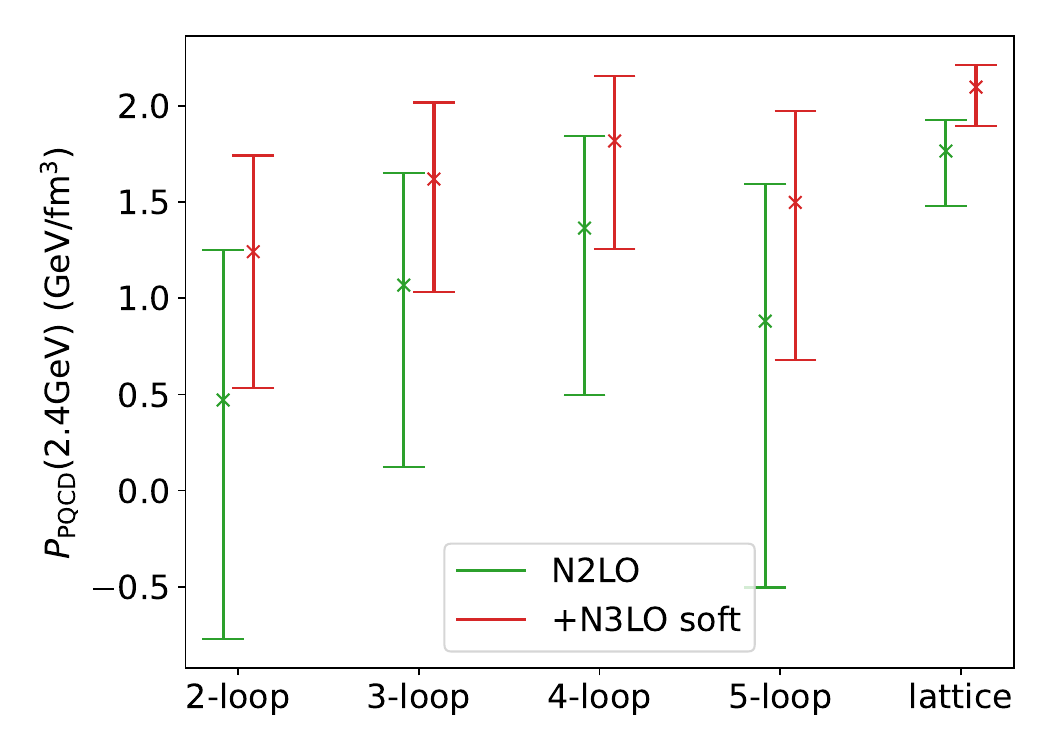}
	\caption{Pressure of the pQCD EOS with $X=1$ at $\mu_B=2.4$ GeV obtained with 2-, 3-, 4-, and 5-loop QCD beta functions. The central values and error bars correspond to those of the reference value $\alpha_s(\bar{\Lambda}=2~\GeV)=0.2994^{+0.0152}_{-0.0141}$. 
	The last column labeled ``lattice'' uses the extracted Landau pole from lattice data~\cite{Bazavov:2014soa} and is obtained with a 4-loop running (see main text). 
	The 2-loop $\alpha_s$ adopted in the main text leads to the largest uncertainties, but is consistent with the N2LO pQCD calculation \cref{eq:pqcd}.
	Notice that the effect of the N3LO soft contribution can be significant, and in most cases appear to improve the convergence of pQCD.
	}
	 \label{fig:p24_al2gev}
\end{figure}

The Landau pole $\bar{\Lambda}_{\msbar}$ is determined once the strong coupling at a given reference scale is specified.
Above, I took the value $\alpha_s(\bar{\Lambda}=2~\GeV)=0.2994^{+0.0152}_{-0.0141}$ from the 2008 particle data group (PDG) report. 
The resulting $P_\pqcd$ and its uncertainties at $\mu_B=2.4$ GeV are shown in \cref{fig:p24_al2gev}.
Since then, PDG only reports $\alpha_s(\bar{\Lambda}=M_Z)$.
Running $\alpha_s$ from the Z boson mass $M_Z\sim100\GeV$ to GeV scales is not straightforward as the decoupling of both charm and bottom quarks occurs in this regime.
But because  the uncertainties of $\alpha_s(M_Z)$  are barely reduced since 2008, one would not expect significant improvements for the bound on $\alpha_s(2~\GeV)$ quoted above.
Furthermore, the current inferred values of $\alpha_s(M_Z)$ could suffer from non-negligible modeling uncertainties as a few subsets of measurements behind the PDG averaged value appear to be in disagreement~\cite{ParticleDataGroup:2022pth}.
A careful running of $\alpha_s$ from $M_Z$ where $N_f=5$ to $\GeV$ scales where $N_f=3$ will be performed in a future work.

An alternative approach is to take the values of Landau pole $\bar\Lambda_{\msbar}$ reported in the literature. This is generally disfavored as these values can be quite sensitive to the order of renormalization group equations used in the analysis.
For instance, in \cref{fig:p24_al2gev} the underlying $\bar{\Lambda}_{\msbar}$ ranges from $\approx310$ to $\approx390$ MeV.
In \cite{Bazavov:2014soa} the authors matched lattice data for the static $q\bar{q}$ potential to a resummed next-to-next-to-next-leading log perturbative calculation in the perturbative regime $\simeq5~\GeV$ and obtained $\bar\Lambda_{\msbar}=315^{+18}_{-12}~\MeV$.
pQCD EOS based on this value is shown as the last column in \cref{fig:p24_al2gev} and is labeled as ``lattice'', where the 4-loop running $\alpha_s$ used throughout \cite{Bazavov:2014soa} is assumed.
It is not clear if some of the choices are better than others.

\section{the partial N3LO pQCD EOS}\label{sec:n3lo}

\begin{figure}
	\includegraphics[width=0.92\linewidth]{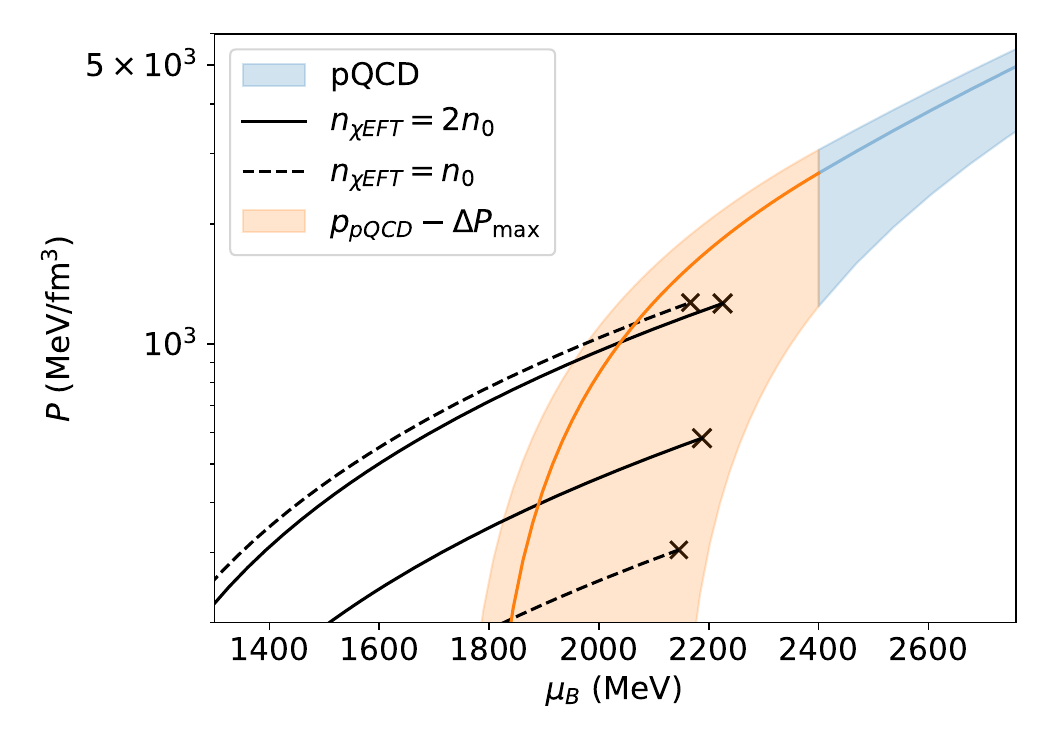}
	\caption{Similar to \cref{fig:dpmax} except that the partial N3LO contribution is included in the pQCD EOS.
	None of the NS EOSs are completely ruled out.
	 }\label{fig:dpmax_n3lo}
\end{figure}

\begin{figure}
	\includegraphics[width=0.92\linewidth]{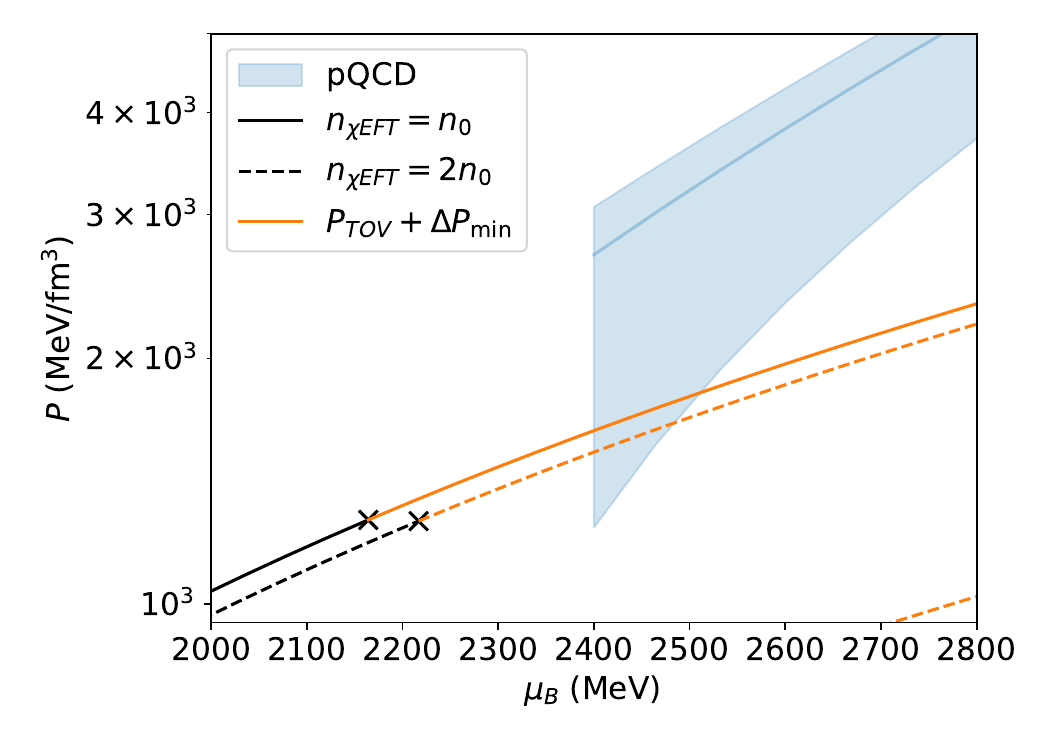}
	\caption{Similar to \cref{fig:dpmin} except that the partial N3LO contribution is included. The effect of \cref{eq:dpmin} becomes weaker. As explained earlier this is not a constraint on NS EOSs. Since predictions for $P_\pqcd$ at given $X$ are not expected to be invariant when higher order terms are included in pQCD, and the uncertainties in $n_\pqcd$ are relatively controlled, it is more convenient to directly parameterize pQCD uncertainties in terms of $P_\pqcd$.
	For instance, if $P_\pqcd(2.4 \GeV)=1.2$ GeV/fm$^3$ and the maximally soft EOSs are confirmed by astrophysics, assuming the ground state is a CFL phase the pairing gap needs to be at least $\simeq100~\MeV$, see \cref{fig:mingap}.
	 }\label{fig:dpmin_n3lo}
\end{figure}

The soft N3LO contribution to the pQCD EOS recently reported in \cite{Gorda:2021kme,Gorda:2018gpy} pushes the pressure higher than that predicted by the N2LO result \cref{eq:pqcd}, as can be seen in \cref{fig:p24_al2gev}.
Although this makes \cref{eq:pmaxbound} more constraining, the main conclusions remain unchanged. 
For instance, as is shown in  \cref{fig:dpmax_n3lo}, the maximally stiff EOSs are still compatible with pQCD predictions assuming $X=1$.
Since the maximally stiff inner core EOS predicts the lowest pressure at $\mu_B\approx 2.2\GeV$, no other valid NS EOSs can be ruled out either. 
As for the $\dpmin$ bound, the increased $P_\pqcd$ due to soft modes at N3LO renders \cref{eq:pminbound} less constraining. And when \cref{eq:pminbound} is violated the tension becomes weaker, see \cref{fig:dpmin_n3lo}.

\section{incorporating constraints on NS sizes}\label{sec:astro}
\begin{figure}[!htbp]
	\includegraphics[width=0.92\linewidth]{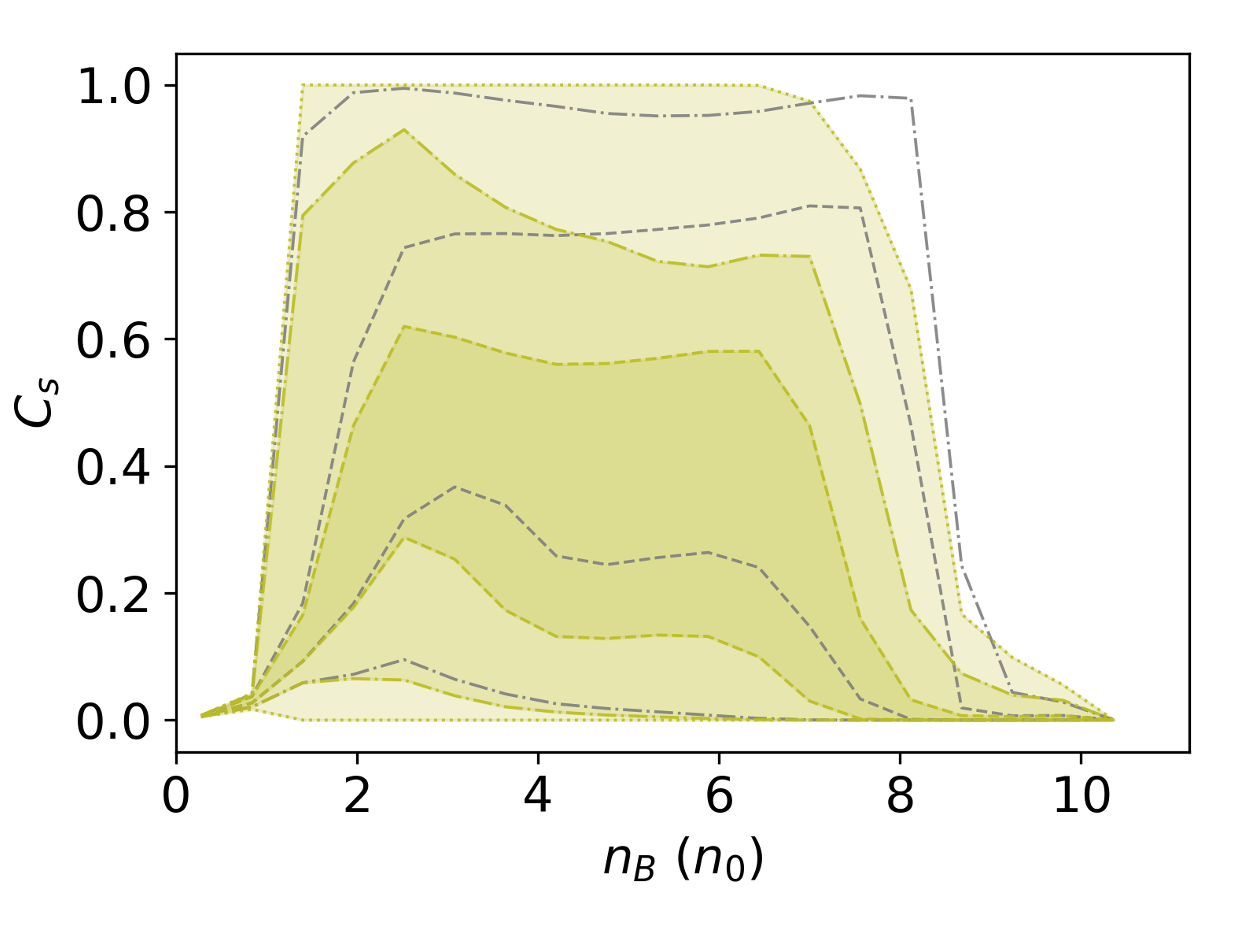}
	\caption{Similar to \cref{fig:muCs} except plotted in the $n_B-C_s$ plane.
	Notice that the prior CIs shown in black are centered around $0.5$ and mostly flat, confirming the peculiar shape seen in \cref{fig:muCs} is a feature of solutions to the TOV equation instead of an indication of bias. Since the pQCD constraint directly impacts $C_s(\mu_B)$, its effects here can be somewhat sensitive to the assumptions underlying the NS EOSs. The results shown here, which drive the CIs lower without introducing additional features, appear to be general and insensitive to those assumptions.
	 }\label{fig:CsN}
\end{figure}

\begin{figure}[!htbp]
	\includegraphics[width=0.92\linewidth]{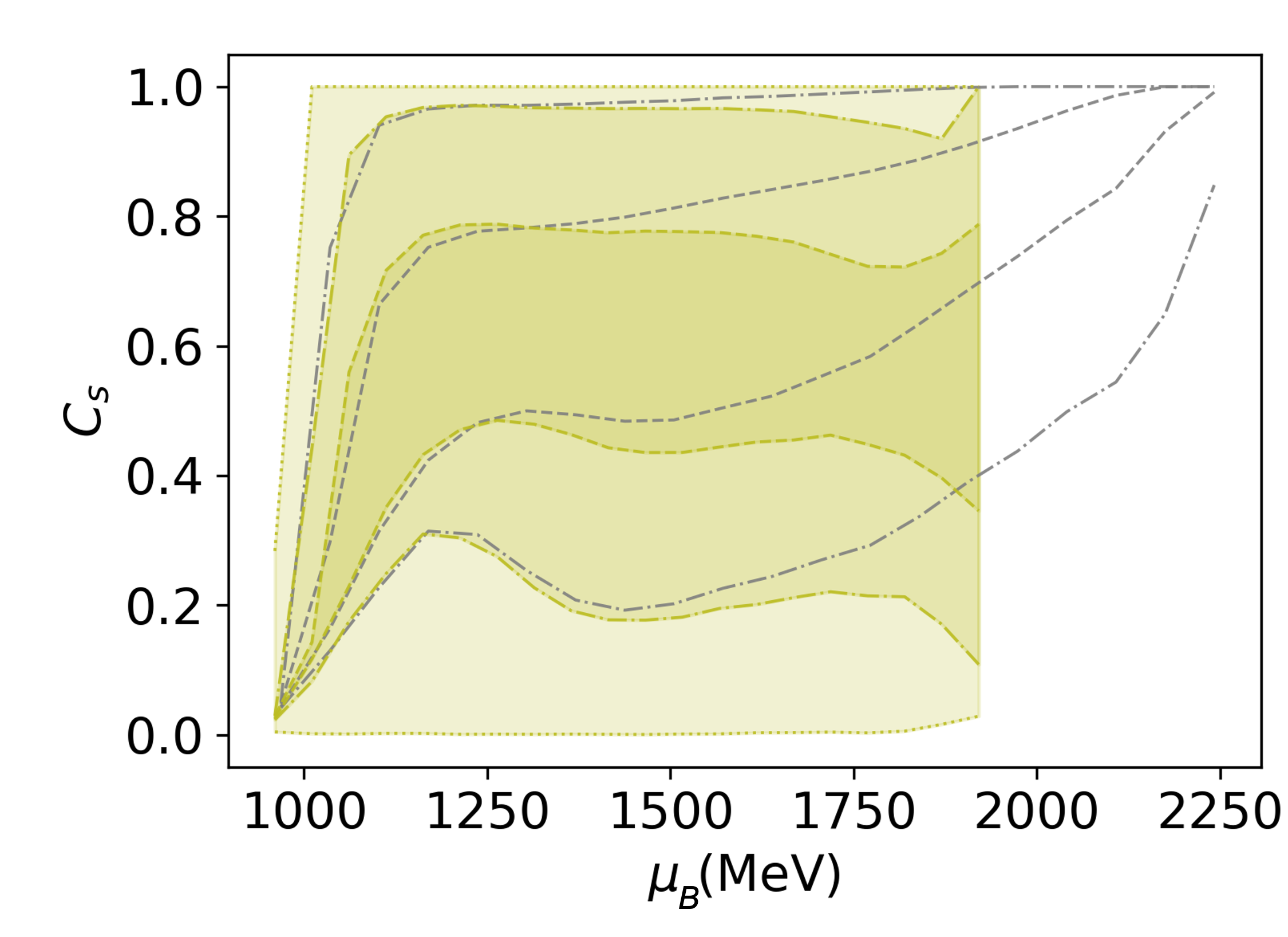}
	\caption{Similar to \cref{fig:muCs} but assumes a prior that already incorporates astrophysical constraints $\Lambda_{1.4\msol}\leq500$ and $M_\tov\leq 2.3 \msol$. When compared to \cref{fig:muCs}, one finds that the astrophysical inputs have negligible effects on $C_s(\mu_B)$, even in this hard cutoff approach ignoring astrophysical uncertainties. 
	The pQCD bound once again shifts 
	the CIs on $C_s$ towards lower values, and forbids $\mu_\tov$ above $\approx1.9$ GeV.
	}\label{fig:CsMuAstro}
\end{figure}

It has been shown in the main text using both the limiting EOSs and randomly generated samples that (potential) pQCD bounds are orthogonal  to those from astrophysical observations. 
The present section aims to strengthen this statement by explicitly taking into account astrophysical constraints.
Specifically, I impose bounds on the tidal deformability of $1.4\msol$ stars $\Lambda_{1.4\msol}\leq500$ from GW170817~\cite{Abbott:2018exr,De:2018uhw,Capano:2019eae}, and the putative upper limit $M_\tov\lesssim2.3M_\odot$ based on speculations that the remnant of GW170817 collapsed to a black hole within seconds~\cite{Margalit:2017dij,Shibata:2019ctb}. 
That the former is orthogonal to pQCD considerations is demonstrated earlier in \cref{fig:l14r20}, and that the latter may interfere with pQCD constraints is mainly due to the closer proximity of the TOV limit to $2\msol$ stars.
All things considered, no appreciable shifts in the observables of $2\msol$ stars are found.
The same observation holds for the posterior CIs on $C_s(\mu_B)$ (see \cref{fig:CsN,fig:CsMuAstro}) where the effects of pQCD considerations appears to be the strongest.

\section{the number density constraint}\label{sec:intcnstr}

\begin{figure}[tbp]
	\includegraphics[width=0.8\linewidth]{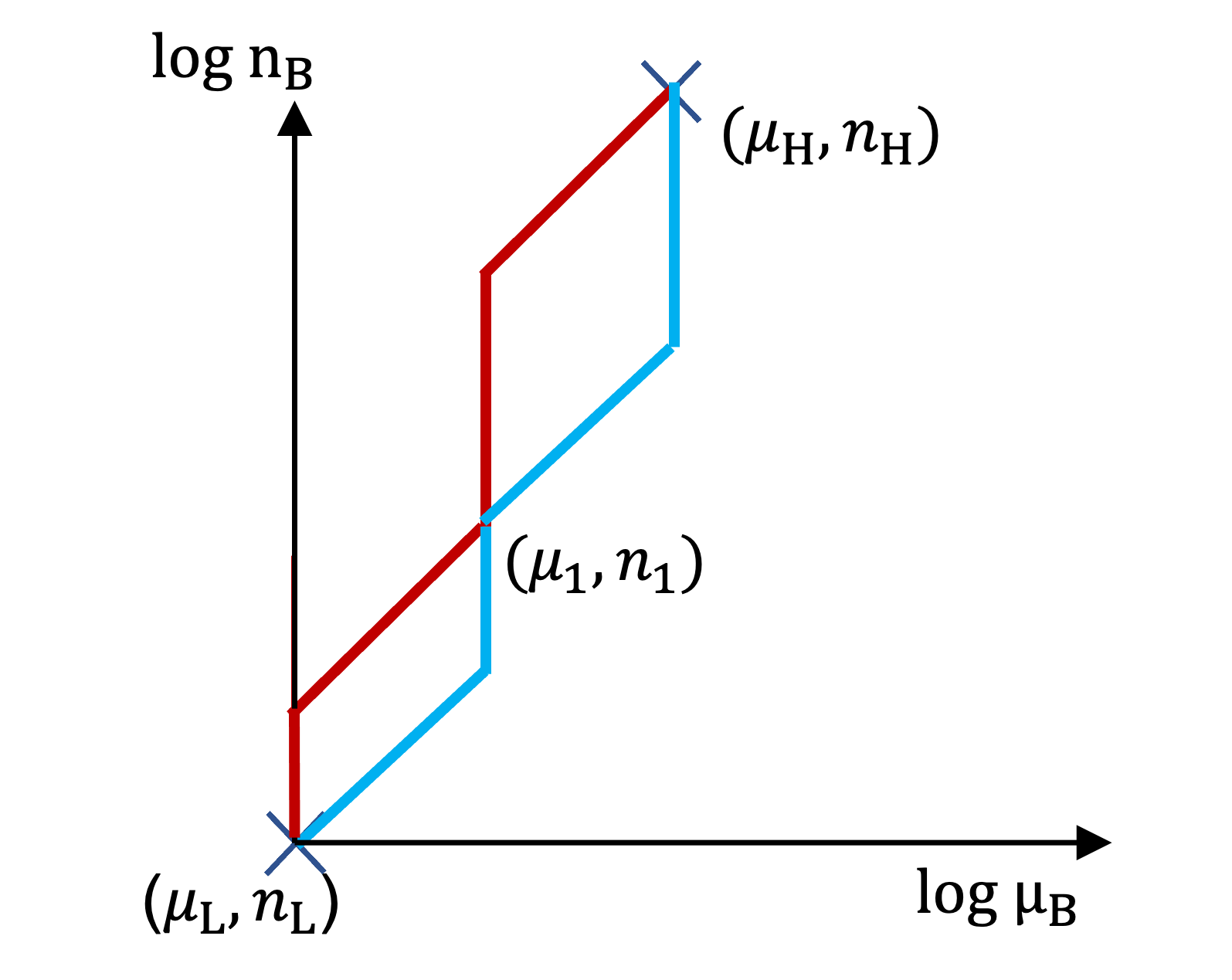}
	\caption{Schematics for the number density constraint for fixed endpoints and $\Delta P_0 =P(\mu_H) - P(\mu_L)$.
	The curve in blue (red) yields $\Delta P_\mathrm{min}$  ($\Delta P_\mathrm{max}$) among all possibilities that pass through the point $(\mu_1, n_1)$. 
	These constructions are obtained by stacking two copies of \cref{fig:construction} on top of each other.
	}\label{fig:demo_nconstraint}
\end{figure}

The low-density matching points in the $\Delta P_\mathrm{min}$ and $\Delta P_\mathrm{max}$ bounds are taken to be at or below the TOV points of NS EOSs, which vary from one EOS to another. 
The number density constraint derived in \cite{Komoltsev:2021jzg} is in a sense a global bound that concerns fixed low- and high-density matching points L and H.
The low- and high-density matching points are taken to be $\mu_\ceft$ and $\mu_\pqcd$ in ~\cite{Komoltsev:2021jzg}.
For all EOSs that pass through an arbitrary point $(\mu_1, n_1)$ between the two endpoints there exist a maximum and a minimum value of $\Delta P$.
These limiting cases are shown as the blue and the red curves in \cref{fig:demo_nconstraint}.
By noting the constructions in the density ranges $[\mu_L,\mu_1]$ and $[\mu_1,\mu_H]$ are two separate realizations of \cref{fig:construction},
the expressions for $\Delta P_\mathrm{min}$ and $\Delta P_\mathrm{max}$ can be obtained by replacing the dummy labels in \cref{eq:dpmax,eq:dpmin}, and are
\begin{align}
	\Delta P_\mathrm{max}&=\frac{n_H \mu_H}{2}\lrb{1-\lrp{\frac{\mu_1}{\mu_H}}^2}
	+\frac{n_1 \mu_1}{2}\lrb{1-\lrp{\frac{\mu_L}{\mu_1}}^2}, \label{eq:dpmax_nden} \\
	\Delta P_\mathrm{min}&=\frac{n_L \mu_L}{2}\lrb{\lrp{\frac{\mu_1}{\mu_L}}^2-1}
	+\frac{n_1 \mu_1}{2}\lrb{\lrp{\frac{\mu_H}{\mu_1}}^2-1}. \label{eq:dpmin_nden} 
\end{align}
In order that there exists an EOS passing through $(\mu_1, n_1)$, $\Delta P_0\equiv P_H-P_L$ has to be sandwiched by these limits $\Delta P_\mathrm{min}\leq \Delta P_0 \leq \Delta P_\mathrm{max}$. But as discussed earlier, only the $\Delta P_\mathrm{max}$ bound may be used as a constraint, and it leads to the number density constraint $n_1(\mu_1)\geq n_\mathrm{min}(\mu_1)$ where
\begin{equation}
n_\mathrm{min}(\mu)=\frac{2\Delta P_0 -\mu_H n_H \lrb{1-\lrp{\frac{\mu}{\mu_H}}^2}}{\mu \lrb{1-\lrp{\frac{\mu_L}{\mu}}^2}}.
\end{equation}

This lower limit can be relevant for large $\mu$ close to $\mu_H$, but it blows off near $\mu_L$. In fact, it crosses the lower bound \cref{eq:nHmin} (noting that `H' and `L' are just dummy labels there) set by the maximally stiff EOS (blue curve in \cref{fig:demo_nconstraint})
at 
\begin{equation}\label{eq:muC}
\mu_C=\sqrt{\mu_L\mu_H\frac{\mu_L n_L-\mu_Hn_H+2\Delta P_0}{\mu_H n_L-\mu_L n_H}}.
\end{equation}
Therefore, for baryon chemical potentials below this value $\mu_C$, \cref{eq:nHmin} 
\begin{equation}
n_\mathrm{min}(\mu)=\frac{\mu}{\mu_L}n_L
\end{equation}
is the relevant bound.

\begin{figure}[htbp]
	\includegraphics[width=0.98\linewidth]{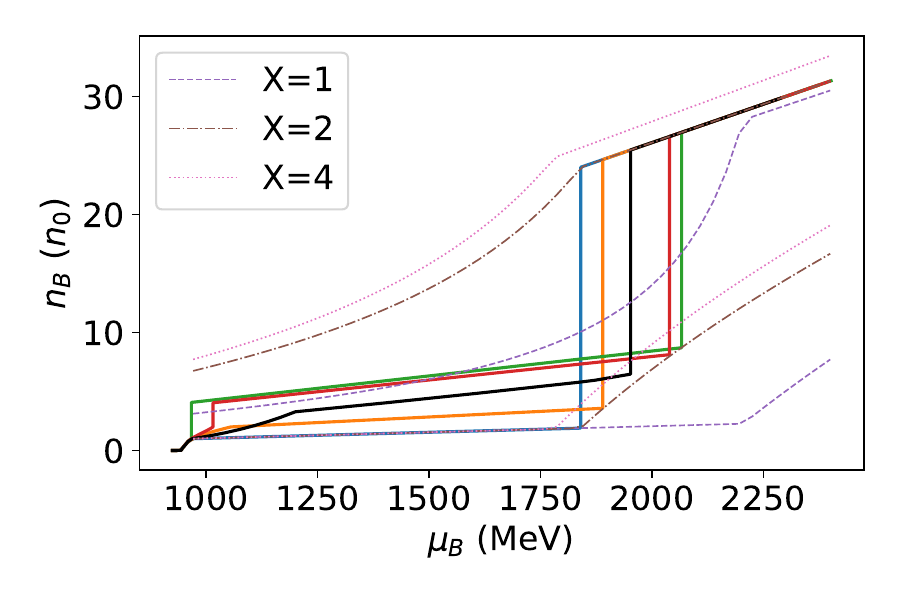}
	\caption{The number density constraints imposed at $\mu_\pqcd=2.4$ GeV for $X=1,2,4$ and at $\mu_L\approx970~\MeV$ (the central value of \ceft~ prediction at $n_0$). 
	Soft modes at N3LO are included in the pQCD EOS.
	The solid colored curves are the modified maximally stiff and maximally soft EOSs compatible with \cref{eq:pmaxbound}. The modification is described in the main text and starts at the critical points shown in \cref{fig:dpmax}. 
	The underlying pQCD parameters are identical to those behind the number density constraints shown here except $X=2$ only.
	The outer curves (green and purple) assume $n_\ceft=n_0$ and the inner ones (red and orange) are for $n_\ceft=2n_0$.
	For the limiting EOSs based on $n_\ceft=n_0$, their critical points sit exactly along the $n_\mathrm{min}$ bound (which assumes the same $n_L=n_0$ and pQCD parameters).
	For other EOSs, for example the solid black line, the critical points do not always reach the $n_\mathrm{min}$ bound, demonstrating that the integral constraint is only a necessary condition.
	}\label{fig:nconstraint}
\end{figure}

The $n_\mathrm{min}$ bound is shown in \cref{fig:nconstraint} as the dashed, dot-dashed, and dotted lines for $X=1,2,4$.
The high-density pQCD point is chosen at $\mu_H=2.4$ GeV, and the N3LO leading log term in the quark matter EOS is also included.
To compare this necessary condition with the one from \cref{eq:pmaxbound}, the limiting NS EOSs modified via the procedure described earlier are shown as the colored solid lines. They are based on the maximally stiff and maximally soft NS EOSs in \cref{fig:dpmax}, and are diverted away at the critical points just before  \cref{eq:pmaxbound} is violated via first-order phase transitions.
For these extreme models, the integral constraint is identical to the $\dpmax$ bound applied at the critical point.
This equivalence can be understood by noting that 
$\Delta P_\mathrm{max}$ does not depend on $n_L$, and the maximally stiff (soft) EOS depicted in blue (red) between $\mu_L$ and $\mu_1$ are, plainly, the maximally stiff (soft) inner core EOS shown in \cref{fig:nconstraint}.
In other words, causal and stable NS EOSs automatically satisfy the contributions to $\Delta P_\mathrm{max}$ below $\mu_1$.
This effectively reduces the problem formulated on the interval $[\mu_L, \mu_H]$ to that on $[\mu_1, \mu_H]$, i.e., \cref{fig:construction}.
For other EOSs that are not maximally stiff or maximally soft,
the extreme construction below $\mu_1$ in \cref{fig:demo_nconstraint} suggests that the integral constraint is weaker than the $\dpmax$ bound.
For instance, in \cref{fig:nconstraint} the solid black curve is required by the $\dpmax$ bound to bend {\em before} it could violate the integral constraint.

For the sake of completeness, I also provide the expression for $n_\mathrm{max}$  derived from 
$\Delta P_\mathrm{min}\leq \Delta P_0$:
\begin{equation}
n_\mathrm{max}(\mu)=\frac{2\Delta P_0 -\mu_L n_L \lrb{\lrp{\frac{\mu}{\mu_L}}^2-1}}{\mu \lrb{\lrp{\frac{\mu_H}{\mu}}^2}-1}.
\end{equation}
It crosses \cref{eq:nLmax} at $\mu_C$ given by \cref{eq:muC}, above which it is replaced by
$$
n_\mathrm{max}(\mu)=\frac{\mu}{\mu_H}n_H.
$$

The $n_\mathrm{max}$ bound is also shown \cref{fig:nconstraint} and appears in the upper region.
It suggests the maximally soft NS EOSs are incompatible with pQCD predictions assuming $X=1$, and the tension surfaces early on at $n_\ceft$ where first-order phase transitions take place.
On one hand, identifying the earliest violation of \cref{eq:pminbound} is a strength of $n_\mathrm{max}$. 
On the other hand, for an intuitive view of the non-perturbative effects required to satisfy \cref{eq:pminbound} (or equivalently the $n_\mathrm{max}$ bound), \cref{fig:dpmin,fig:dpmin_n3lo} are better suited.

\section{Model-independent upper bound on $\Delta_\mathrm{CFL}$}\label{sec:maxgap}

Here, I report model-independent bounds on the highest $\Delta_\cfl$ allowed by given NS maximum masses.
These constraints improve upon the results reported in ref \cite{Kurkela:2024xfh} in that they do not require any assumption about NS inner cores.

Taking into account pairing contributions to the EOS \cref{eq:cfleos}, the $\dpmax$ bound ~\cref{eq:pmaxbound} becomes
\begin{equation}\label{eq:maxgap0}
P_\pqcd+P_\cfl-P_\tov \leq\dpmax
\end{equation}
from which one obtains
\begin{equation}\label{eq:maxgap}
P_\cfl\leq \dpmax-\Delta P.
\end{equation}
The value of $\Delta_\cfl$ can be solved from $P_\cfl$ via \cref{eq:pcfl}.
For clarity, with the pairing effects included the quantities above are 
$$\dpmax=\frac{n_\cqm \mu_\cqm}{2}\lrb{1-\lrp{\frac{\mu_\tov}{\mu_\cqm}}^2}$$
and
$$\Delta P=P_\pqcd(\mu_\cqm)-P_\tov$$
where $\mu_\cqm\equiv\mu_\pqcd\equiv\mu_H$ is the baryon chemical potential at the chosen high-density matching point, $n_\cqm$ is given by \cref{eq:cfleos}, and $P_\pqcd(\mu_\cqm)$ is the perturbative contribution to the pressure.

\begin{figure}[!htbp]
	\includegraphics[width=0.98\linewidth]{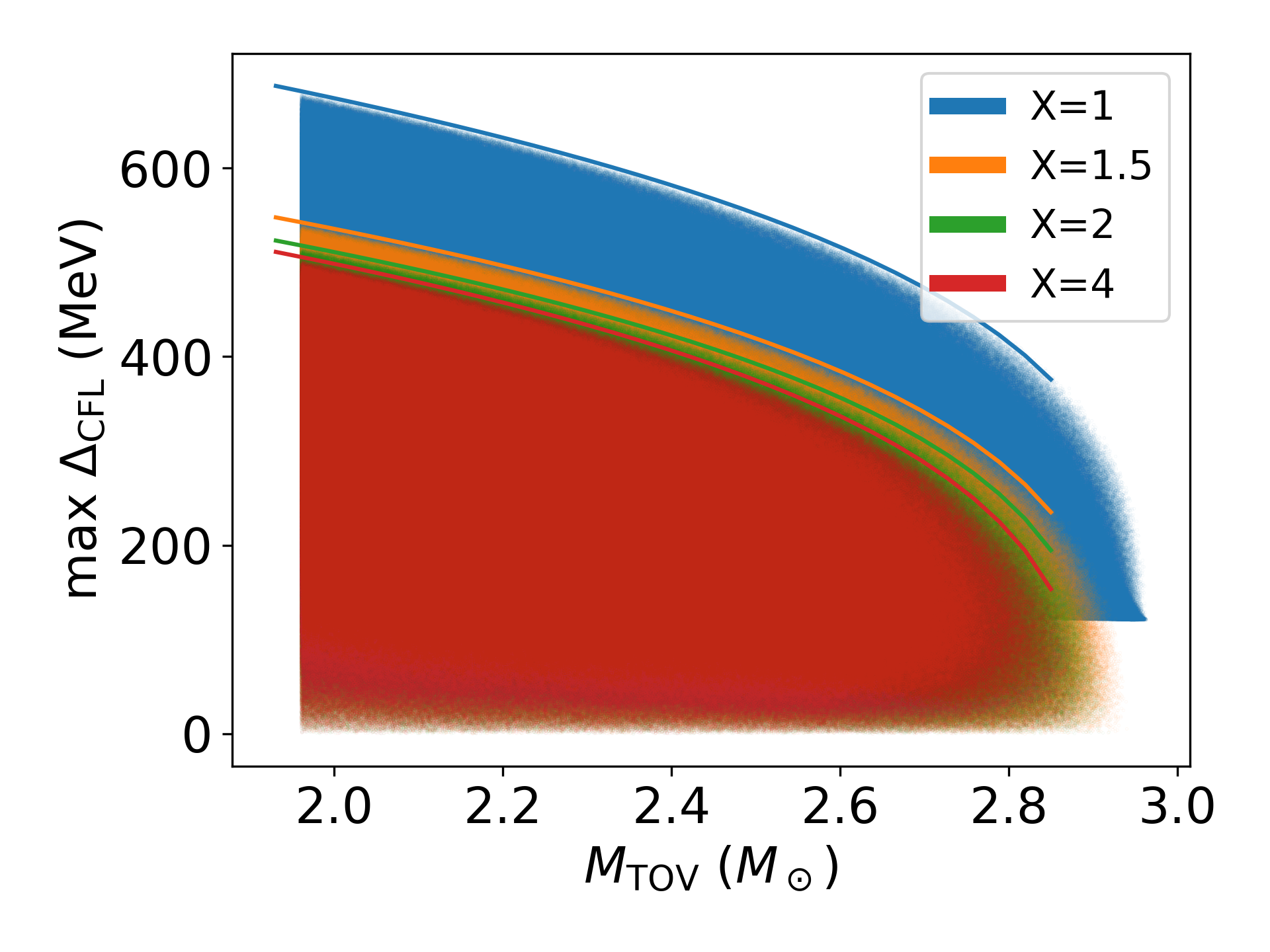}
	\caption{Maximum allowed CFL pairing gaps as a function of $M_\tov$. 
	The dots are computed for each EOS sample by solving $\Delta_\cfl$ that saturates the inequality \cref{eq:maxgap}.
	The lines are based on the maximally soft NS EOSs and give the global upper bound over all NS inner core EOSs.
	The soft N3LO pQCD contribution to the CQM EOS is included.
	  }\label{fig:maxgap}
\end{figure}

For chosen pQCD EOS and the high-density matching point, the RHS of \cref{eq:maxgap} is a function of the TOV point and gives an upper bound $\max\{\Delta_\cfl\}$ when the equality holds.
It has a global maximum $\mathbf{max}\{\Delta_\cfl\}\equiv \max_\mathrm{NS}\{\max\{\Delta_\cfl\}\}$ across all NS EOSs reached by the maximally soft NS inner core EOSs.
This is because the maximally soft NS EOSs predict the highest $P_\tov$ at the highest $\mu_\tov$.
Large $\Delta_\cfl$ yielding $P_\cqm$ too high for the maximally soft NS EOS will certainly overmatch other NS EOSs as well.
This upper bound is shown in \cref{fig:maxgap} along with sample-based statistics.
The constraints strengthen rapidly with increasing $X$ near $X\approx1$, largely due to the $\Delta_\cfl^2$ dependence in $P_\cfl$.
In the absence of additional assumptions, the CFL pairing gap at $\mu_B=2.4~\GeV$ can be robustly placed below $\simeq500~\MeV$ by the existence of two-solar-mass pulsars~\cite{Demorest:2010bx,Antoniadis:2013pzd,NANOGrav:2019jur,Romani:2021xmb,Fonseca:2021wxt}.

\clearpage
\newpage

%


\end{document}